\documentclass[iop,apj]{emulateapj}

\usepackage[caption=false]{subfig}

\usepackage{times}

\usepackage{amsmath, amssymb, amstext}
\usepackage[breaklinks, colorlinks, citecolor=blue, linkcolor=magenta]{hyperref} 

\usepackage[all]{hypcap} 

\shorttitle{The ELQS-S Sample and the QLF}

\shortauthors{Schindler et al.}

\begin{document}

\title{The Extremely Luminous Quasar Survey in the Sloan Digital Sky Survey footprint. III. The South Galactic Cap Sample and the Quasar Luminosity Function at Cosmic Noon}

\author{Jan-Torge Schindler\altaffilmark{1}\altaffilmark{2}}
\author{Xiaohui Fan\altaffilmark{1}}
\author{Ian D. McGreer\altaffilmark{1}}
\author{Jinyi Yang\altaffilmark{1}}
\author{Feige Wang\altaffilmark{3}}
\author{Richard Green\altaffilmark{1}}

\author{Johan P. U. Fynbo\altaffilmark{4}}
\author{Jens-Kristian Krogager\altaffilmark{5}}

\author{Elisabeth M. Green\altaffilmark{1}}
\author{Yun-Hsin Huang\altaffilmark{1}}
\author{Jennifer Kadowaki\altaffilmark{1}}
\author{Anna Patej\altaffilmark{1}}
\author{Ya-Lin Wu\altaffilmark{1}\altaffilmark{6}\altaffilmark{7}}
\author{Minghao Yue\altaffilmark{1}}

\altaffiltext{1}{Steward Observatory, University of Arizona, 933 North Cherry Avenue, Tucson, AZ 85721, USA}
\altaffiltext{2}{Max-Planck Institute for Astronomy, Königstuhl 17, 69117 Heidelberg, Germany}
\altaffiltext{3}{Department of Physics, University of California Santa Barbara,  Santa Barbara, CA 93106-9530}
\altaffiltext{4}{Cosmic Dawn Center (DAWN), Niels Bohr Institute, University of Copenhagen, Juliane Maries Vej 30, DK-2100 Copenhagen $\O$; DTU-Space, Technical University of Denmark,  Elektrovej 327, DK-2800 Kgs. Lyngby, Denmark}
\altaffiltext{5}{Institut d'Astrophysique de Paris, CNRS-UPMC, UMR7095, 98bis bd Arago, F-75014 Paris, France}
\altaffiltext{6}{Department of Astronomy, University of Texas at Austin, 2515 Speedway, Stop C1400 Austin, Texas 78712-1205, USA}
\altaffilmark{7}{51 Pegasi b Fellow}
\defcitealias{Schindler2017}{Paper I}
\defcitealias{Schindler2018a}{Paper II}

\begin{abstract}

 We have designed the Extremely Luminous Quasar Survey (ELQS) to provide a highly complete census of unobscured UV-bright quasars during the cosmic noon, $z=2.8-5.0$. 
 Here we report the discovery of 70 new quasars in the ELQS South Galactic Cap (ELQS-S) quasar sample, doubling the number of known extremely luminous quasars in $4,237.3\,\rm{deg}^2$ of the SDSS footprint. 
 These observations conclude the ELQS and we present the properties of the full ELQS quasar catalog, containing 407 quasars over $11,838.5\,\rm{deg}^2$. Our novel ELQS quasar selection strategy resulted in unprecedented completeness at the bright end and allowed us to discover 109 new quasars in total. This marks an increase of $\sim36\%$ (109/298) to the known population at these redshifts and magnitudes, while we further are able to retain a selection efficiency of $\sim80\%$.
 On the basis of 166 quasars from the full ELQS quasar catalog, who adhere to the uniform criteria of the 2MASS point source catalog, we measure the bright-end quasar luminosity function (QLF) and extend it one magnitude brighter than previous studies.
 Assuming a single power law with exponential density evolution for the functional form of the QLF, we retrieve the best fit parameters from a maximum likelihood analysis. We find a steep bright-end slope of $\beta\approx-4.1$ and we can constrain the bright-end slope to $\beta\leq-3.4$ with $99\%$ confidence. The density is well modeled by the exponential redshift evolution, resulting in a moderate decrease with redshift ($\gamma\approx-0.4$).  
\end{abstract}




\section{Introduction}


Quasars are the most luminous, non-transient light sources in the universe. 
Their strong emission emanates from the accretion disk around rapidly growing supermassive black holes (SMBH) at the centers of galaxies. 
The study of quasars provides crucial insight into the formation and evolution of galaxies as the mass of the SMBH and properties of the host galaxy show strong correlations \citep[see][for a review]{Kormendy2013}.
Quasars discovered within the first billion years of the universe \citep{Mortlock2011, Banados2018} probe the era of reionization and place strong constraints on the formation and growth of SMBHs.
As bright background sources, they have also furthered our understanding of the nature and evolution of the intervening intergalactic galactic medium \citep{Simcoe2004, Prochaska2005, Worseck2011}.


Our understanding of the cosmic growth of SMBHs strongly relies on the demographics of the quasar population with the quasar luminosity function (QLF) being one of the most fundamental probes. The QLF is best described by a broken double power law \citep{Boyle1988, Boyle2000, Pei1995}, characterized by a faint-end slope, a bright-end slope, an overall normalization and a break luminosity, where the slopes change. The faint-end slope is generally flatter than the bright-end slope and all four parameters possibly change with redshift.

Large volume spectroscopic surveys, like the Sloan Digital Sky Survey \citep[SDSS;][]{York2000}, the Baryon Oscillation Spectroscopic Survey (BOSS; \citet{Eisenstein2011, Dawson2013}) and the extended BOSS (eBOSS; \citet{Dawson2016, Blanton2017}), built the largest optical quasar sample to date. It allowed to tightly constrain the QLF of UV-bright unobscured quasars over a wide range of luminosities and redshifts ($0.3\lesssim z \lesssim 5$). At higher redshifts ($z\geq5$) specifically targeted surveys have constrained the QLF \citep[e.g.][]{Jiang2008, McGreer2013, Yang2016}.

At intermediate redshifts ($z=2.8-4.5$) there has been a standing debate in the literature on the evolution of the bright-end slope. Some earlier studies suggested that the bright-end slope would flatten with redshift \citep{Koo1988, Schmidt1995, Fan2001b, Richards2006}. However, more recent estimates of the QLF seem to indicate that the bright-end slope remains steep up to the highest redshifts \citep{Jiang2008, Croom2009, Willott2010a,  McGreer2013, Yang2016}

This is the third paper in a series presenting the Extremely Luminous Quasars Survey (ELQS), a spectroscopic survey focused on the bright end ($m_{\rm{i}}\leq18.0$, $M_{1450}<-27$) of the UV-bright type-I quasar distribution at $z\geq2.8$.

The first paper \citep[][hereafter Paper\,I]{Schindler2017} discussed the incompleteness of the SDSS spectroscopic quasar survey and BOSS for very bright quasars at these redshifts and showcases our novel quasar selection method.

In the second paper of this series \citep[][hereafter Paper\,II]{Schindler2018a} we presented the ELQS quasar sample in the North Galactic Cap (ELQS-N; $90\,\rm{deg} {<} \rm{RA} {<} 270\,\rm{deg}$) and a first estimate of the bright-end QLF.

This work presents the final ELQS quasar catalog, covering the entire SDSS footprint ($11,838.5\pm20.1\,\rm{deg}^2$), and the resulting QLF at the bright end at redshifts $2.8\leq z\leq4.5$. We also report the results of our spectroscopic identification campaign in the South Galactic Cap, the ELQS-S sample. 
We provide a brief introduction to the ELQS survey in Section\,\ref{sec_ELQS_intro}. Subsequently, we discuss the ELQS-S observations and our data reduction in Section\,\ref{sec_obervations}. The ELQS-S sample, including the discovery of 70 new quasars, is presented in Section\,\ref{sec_ELQS-S}, which leads to a discussion of the properties of the full ELQS quasar catalog (Section\,\ref{sec_ELQS_QC}). Based on this catalog we calculate the QLF (Section\,\ref{sec_QLF}) and discuss the implications of our results in Section\,\ref{sec_discussion}. We summarize our findings in Section\,\ref{sec_conclusion}.

All magnitudes are displayed in the AB system \citep{Oke1983} and corrected for galactic extinction \citep{Schlafly2011} unless otherwise noted.
We denote magnitudes not corrected for galactic extinction only by $x$, where $x$ refers to the wavelength band in question, as opposed to extinction corrected magnitudes $m_{\rm{x}}$.
We adopt the standard flat $\Lambda\rm{CDM}$ cosmology with $H_0=70\,\rm{km}\rm{s}^{-1}\rm{Mpc}^{-1}$, $\Omega_{\rm{m}}=0.3$ and $\Omega_\Lambda=0.7$ in general consistent with recent measurements \citep{PlanckCollaboration2016}.

\section{Introduction to the Extremely Luminous Quasar Survey}\label{sec_ELQS_intro}


The Extremely Luminous Quasar Survey (ELQS) was designed to provide an accurate measure of the UV-bright type-I QLF at the bright end ($M_{1450}<-27$) at intermediate redshifts ($2.8 \leq z \leq 4.5$). 

We apply a highly inclusive color cut in the J-K-W2 plane ($ \rm{K}-\rm{W2} \ge 1.8 - 0.848 \cdot \left(\rm{J}-\rm{K} \right)$; Vega magnitudes) using photometry from the Two Micron All Sky Survey  \citep[2MASS;][]{Skrutskie2006} and the Wide-field Infrared Survey Explorer mission \citep[WISE;][]{Wright2010}. 
Using optical SDSS photometry along with WISE photometry we estimate photometric redshifts and further classify our candidates using random forests \citep{Breiman2001}, a supervised machine learning technique. In both cases the random forest method is trained on a quasar sample built from the SDSS DR7 and DR12 quasar catalogs \citep{Schneider2010, Paris2017}. The quasar selection is described in \citetalias{Schindler2017}.

The ELQS covers the entirety of the SDSS footprint excluding the galactic plane ($b<-20$ or $b>30$),  but including the SDSS strips at $\rm{Decl.}<0$ .We have estimated the area coverage of our survey in \citetalias{Schindler2017} using the Hierarchical Equal Area isoLatitude Pixelation \citep[HEALPix;][]{Gorski2005}. A description of the calculation process and the general parameters used can be found in \citet{Jiang2016}. The effective area of the ELQS is $11,838.5\pm20.1\,\rm{deg}^2$, of which  $7,601.2\pm7.2\,\rm{deg}^2$ are part of the North Galactic Cap ($90\,\rm{deg} {<} \rm{RA} {<} 270\,\rm{deg}$) and $4,237.3\pm12.9\,\rm{deg}^2$ are part of the South Galactic Cap ($\rm{RA} {>} 270\,\rm{deg}$ and $\rm{RA} {<} 90\,\rm{deg}$).

\begin{figure*}
 \centering 
 \includegraphics[width=\textwidth]{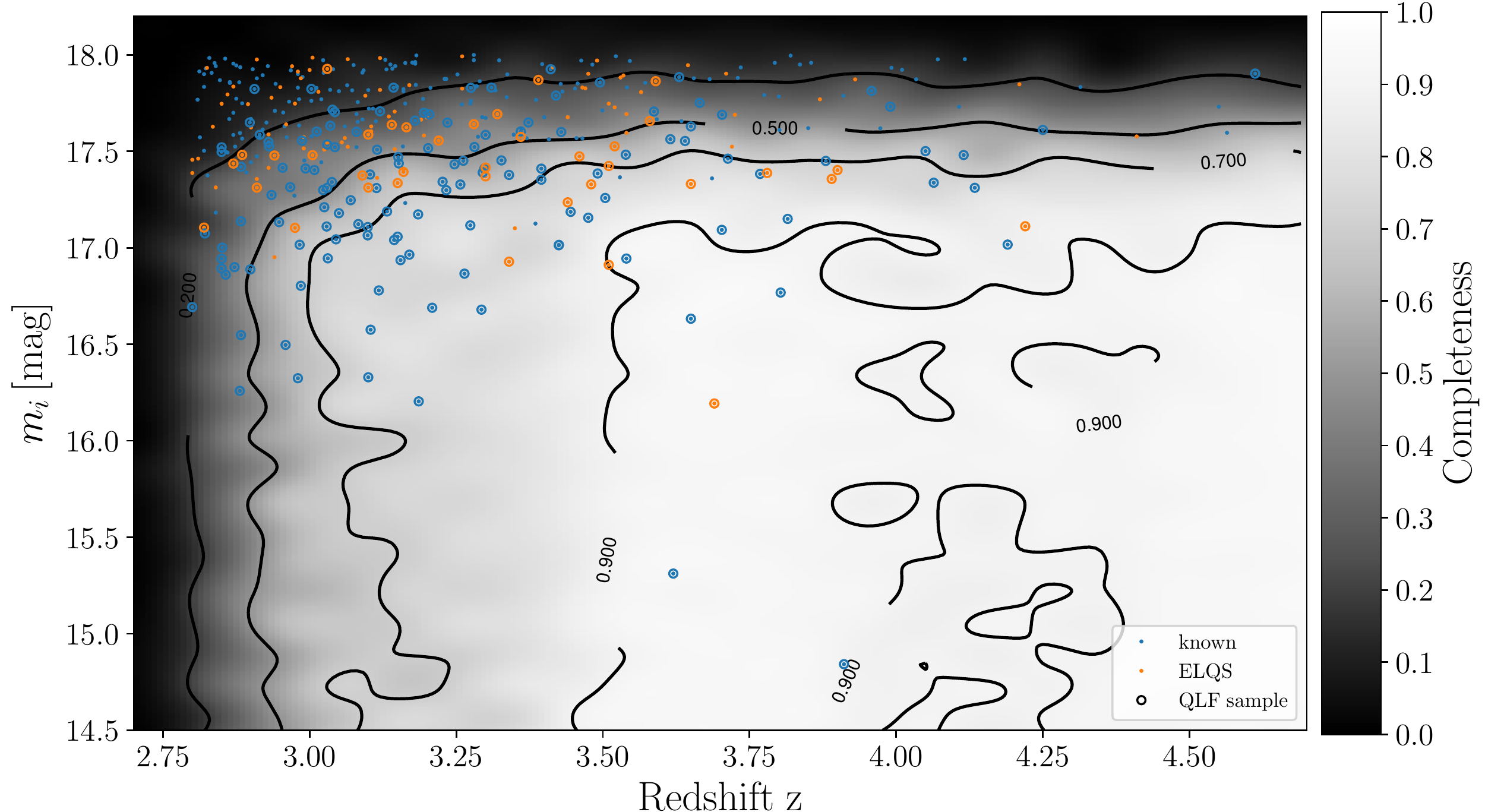}
 \caption{The selection function (completeness) of the ELQS as a function of redshift and $i$-band magnitude. Contour levels are drawn with solid lines at $20\%$, $50\%$, $70\%$ and $90\%$. Newly discovered and already known quasars in the full ELQS quasar sample are displayed in orange and blue, respectively. All ELQS quasars, which are part of the QLF sample are highlighted.}
 \label{fig_completeness}
\end{figure*}

We discuss our selection function in the second paper of this series \citepalias{Schindler2018a}. For its calculation, we imposed our selection criteria, including the completeness limits of the photometric catalogs, on a sample of simulated quasar spectra uniformly distributed as a function of observed i-band magnitude and redshift. The resulting completeness reaches $>70\%$ in the core region of our survey ($3.0\lesssim z \lesssim 5.0$; $m_{\rm{i}}\lesssim 17.5$). 
We show the selection function of the ELQS survey as a function of redshift and apparent $i$-band magnitude in Figure\,\ref{fig_completeness}, highlighting all newly discovered and already known quasars of the full ELQS sample in orange and blue, respectively.  

We also presented the ELQS quasar sample in the North Galactic Cap ($90\,\rm{deg} \leq RA \leq 270\,\rm{deg}$) footprint (ELQS-N) in \citetalias{Schindler2018a}. This sample consists of 270 quasars at $m_{\rm{i}} \leq 18.0$ and $z\geq2.8$ of which 39 were newly identified as part of the ELQS survey. 

Using 120 quasars from the ELQS-N sample, which adhere to the uniform photometric criteria of the 2MASS point source catalog assumed by our selection function calculation, we conducted a first analysis of the bright-end QLF . Single power law fits to the data result in a steep value for the bright-end slope of $\beta\approx-4$. We further can constrain the bright-end slope to $\beta<-2.94$ with $99\%$ confidence. This result contrasts earlier QLF estimates at the same redshift \citep{Fan2001b, Richards2006}, who find a generally flatter slope of $\beta\approx-2.5$.

The present work completes the ELQS survey with spectroscopic observations in the South Galactic Cap of the SDSS footprint. Our selection for this area resulted in a larger quasar candidate sample than for the ELQS-N, including many quasars that were not spectroscopically followed up by the original SDSS quasar survey. As a consequence, the ELQS-S sample presents a total of 70 newly discovered quasars, which allow for stronger statistical constraints on the QLF.

\subsection{ELQS Candidates in the Literature}

We have discussed the references for known quasars in the ELQS-N in some detail in \citetalias[][Section\,2.2]{Schindler2018a}. Since all known quasars in the ELQS-S sample are from the same references, we will only present a summary below. For further details, please refer to \citetalias{Schindler2018a}.

The majority of known quasars in the ELQS-S sample were discovered by the SDSS \citep{Abazajian2009}, the BOSS and eBOSS. The quasars are published in the  SDSS DR7 \citep{Schneider2010} , DR12 \citep{Paris2017} and DR14 \citep{Paris2018} quasar catalogs.

In addition, we have matched against the Million Quasar Catalog \citep[MQC; ][]{Flesch2015} to identify known quasars, which were not included in the SDSS quasar catalogs. The MQC is a compilation of quasars from a variety of different sources in the literature and includes quasar candidates as well. Only verified quasars were used in the cross-match between the catalog and our candidates.

We also match our quasar candidates against the most recent catalogs of the The Large Sky Area Multi-Object Fibre Spectroscopic Telescope (LAMOST) quasar survey \citep{Dong2018}. The LAMOST quasar survey is part of the LAMOST ExtraGAlactic Survey \citep[LEGAS; ][]{Zhao2012} and quasars are selected using multi-color photometry color cuts as well as data-mining algorithms.
Three candidates of the ELQS-S sample are successfully matches to LAMOST quasars.

Furthermore, Yang et al. (publication in preparation) are currently carrying out a spectroscopic survey similar to the ELQS. Their candidate selection consists of two samples that use optical and infrared color criteria presented in \citet{Wu2010} and \citet{Wu2012}. They aim to find bright quasars at $z\approx2-3$ and at $z\geq4$ missed by the SDSS/BOSS/eBOSS quasar surveys and to test different quasar selection criteria for the upcoming LAMOST quasar survey. Their spectroscopic observations are conducted at the Lijiang telescope ($2.4\,\rm{m}$) and the Xinglong telescope ($2.16\,\rm{m}$).

We also discovered that one of our candidates, J215743.62+233037.1, was part of the HST GO program 13013\footnote{\url{http://www.stsci.edu/hst/phase2-public/13013.pro}} (PI: G. Worseck). While it was never published in a quasar catalog, it has been further studied by \citep{Zheng2015} and \citet{Schmidt2017}. We decided to include it in our sample of newly discovered ELQS-S quasars, to formally publish its classification, including an optical spectrum.

\section{Spectroscopic Observations and Data Reduction}\label{sec_obervations}

\begin{figure}
\center
 \includegraphics[width=0.5\textwidth]{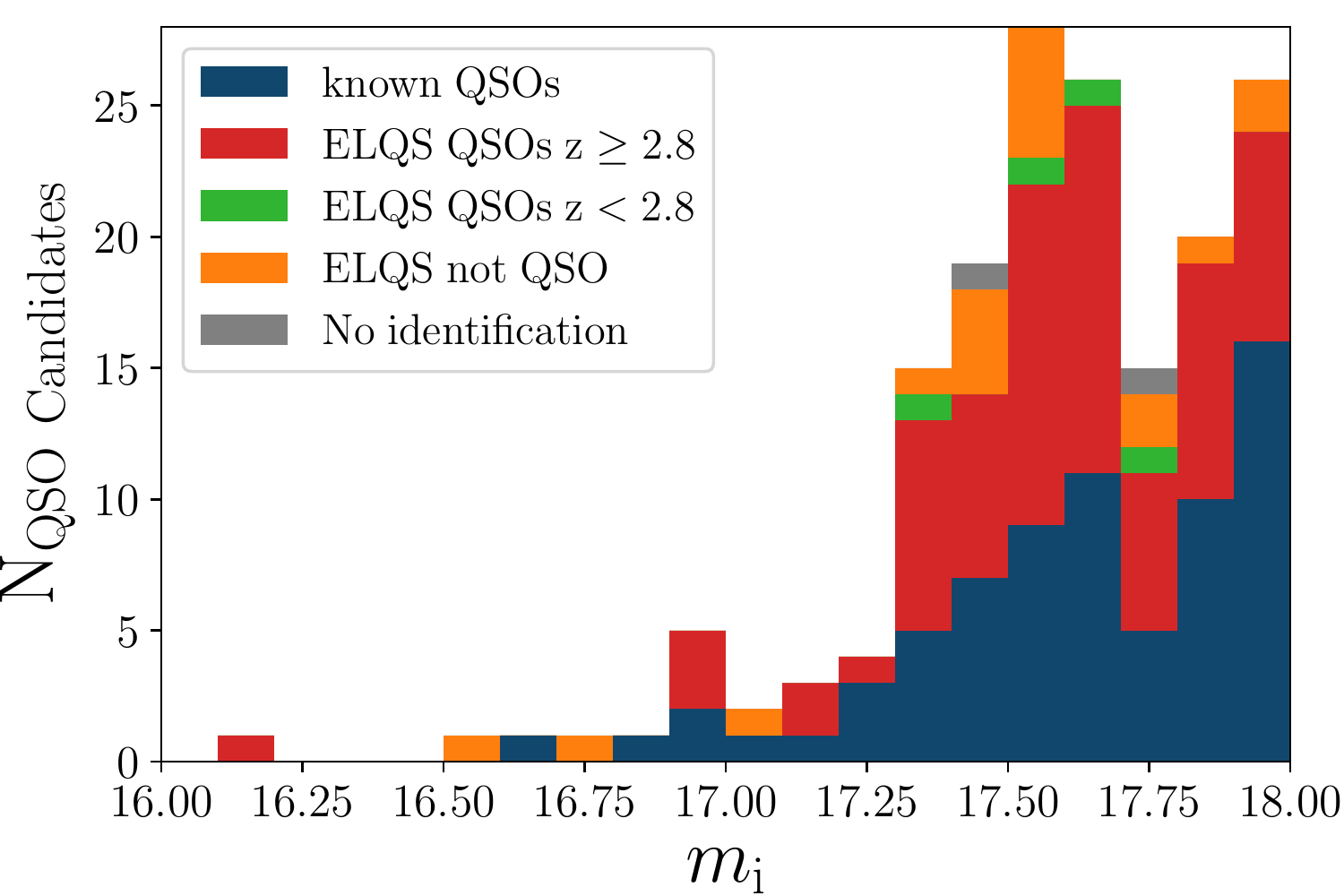}
 \caption{We present the distribution of all good primary ELQS-S candidates as a function of their apparent SDSS i-band magnitude. 
 Quasars known from the literature are colored blue. Red and green colors highlight the newly discovered quasars with $z\geq2.8$ and $z<2.8$, respectively. Candidates that have been identified not to be quasars are shown in orange, while objects that could not be identified or were not observed are shown in gray.}
 \label{fig_elqs_spec_completeness}
\end{figure}

Exploratory observations for the ELQS started in 2015 and were designed to test a variety of selection criteria. As a result we discovered a range of quasars that are not included in the primary ELQS candidate catalog, which was finalized in September 2016. We present their discovery spectra and their general properties in Appendix\,\ref{app_addquasars}.

Observations of the ELQS-S sample have been completed and 96 out of 97 candidates were observed with a range of different telescopes. These include the Vatican Advanced Technology Telescope (VATT), the MMT $6.5\,\rm{m}$ telescope, the $90$-inch ($2.3\,\rm{m}$) Bok Telescope, the Nordic Optical Telescope (NOT) and the ($4.1\,\rm{m}$) Southern Astrophysical Research Telescope (SOAR).
In this section we will detail the different instrumental setups and briefly describe the data reduction process.

\subsection{VATT Observations}
We have carried out the majority of our spectroscopic identifications with the VATTSpec spectrograph on the VATT. We used the  $300\,$g/mm grating in first order blazed at $5000\,\text{\AA}$. The spectra have a resolution of $R\sim1000$ ($1\farcs5$ slit) and a coverage of $\sim4000\,\text{\AA}$ around our chosen central wavelength of $\sim5775\,\text{\AA}$. 

The observations for the ELQS-S were conducted in multiple campaigns. Pilot observations started in 2015 October 8-12. The program continued in 2016 November 20-23 and December 18-20. In 2017 we finished the South Galactic Cap footprint during observations on November 7-12.
Depending on the object and the conditions the exposure times varied between 15 and 30 minutes.

\subsection{Bok Observations}
In fall 2016 we were awarded three nights on the Bok telescope. We used the Boller \& Chivens Spectrograph (B\&C spectrograph) with the 400\,g/mm grating blazed at $4889\AA$ in first order and the UV-36 blocking filter. The central wavelength was chosen to be $\sim5250\AA$, resulting in a coverage of $\approx3655-6850\AA$. The observations were conducted in 2016 on October 13-14 and November 15. The spectra were taken with the $2\farcs5$ slit, resulting in a resolution of $R\approx750$. 
Depending on weather conditions and the apparent magnitude of the object we have used exposure times of $\sim5-15\,\text{min}$.

\subsection{MMT Observations}
We have used the MMT Red Channel Spectrograph to carry out follow-up observations of our newly discovered quasars. For all observations we have used the MMT 270\,g/mm and 300\,g/mm gratings blazed at 1st/$7300\text{\AA}$ and 1st/$4800\text{\AA}$, respectively. With regard to the 270\,g/mm grating we used central wavelengths of $6400\text{\AA}$, $7000\text{\AA}$ and $7150\text{\AA}$. For the 300\,g/mm we used central wavelengths of $5000\text{\AA}$, $5500\text{\AA}$ and $6083\text{\AA}$. The 270\,g/mm grating has an approximate coverage of $3705\text{\AA}$, whereas the 300\,g/mm grating has an approximate coverage of $3310\text{\AA}$. We chose exposure times of $\sim3-15\,\text{min}$ per spectrum, depending on the object and conditions.
Based on the seeing conditions, we have either used the $1\farcs25$ or the $1\farcs5$ slit, providing a resolution of $R\approx300-400$ with both gratings. Observations were taken in 2017 on May 17-18, October 20-21 and November 16 and on January 20 2018. 

After the completion of the survey we noticed that the MMT Red Channel Spectrograph dim continuum lamp failed during our run on May 17-18 2017 resulting in very low signal to noise flat fields for those two nights. We were able to re-reduce the spectra with the 300\,g/mm grating and a central wavelength of  $5560\text{\AA}$ using flat fields from a different observing run. The low signal-to-noise flat fields are still used for all spectra centered around $6083\text{\AA}$, introducing additional noise. However, we do not expect any systematic biases as the detector of the spectrograph  does not show strong sensitivity variations along the spatial direction and variations along the dispersion direction are indirectly taken care of by the standard calibration procedure.

\subsection{Nordic Optical Telescope Observations}
In 2017, some identification spectra were taken during the Nordic Optical Telescope (NOT) summer schools (August 23-25, September 5-9). These observations were conducted with the Andalucia Faint Object Spectrograph and Camera (ALFOSC) using the 300 g/mm grism (\#4). The grism, centered around $5800\text{\AA}$, offers a wavelength coverage of $3200\text{\AA}-9600\text{\AA}$. We used the blue blocking filter WG345 356\_LP with a cut on at $\sim3560\text{\AA}$. Given the above setup the spectra, taken with the $1\farcs0$ and the $1\farcs3$ slit, provide a resolution of $R\approx360$ and $R\approx280$, respectively. Exposure times varied between 2.5 to 5 min, depending on atmospheric transparency and apparent target magnitude.

\subsection{SOAR}
In addition, we observed quasar candidates with the Goodman High Throughput Spectrograph (Goodman HTS) on the Southern Astrophysical Research (SOAR) Telescope ($4.1\,\rm{m}$). These observations were carried out in 2017 October 6-10 and 2018 January 22-24. We used the 400\,g/mm grating with central wavelengths of $6000\AA$ and $7300\AA$. The spectra have a spectral coverage of $\sim4000-8000\AA$ and $\sim5300-9300\AA$, respectively. The first setup used the GG-385 blocking filter, whereas the second one used the GG-495 blocking filter. We used the red camera in 2x2 spectral mode for all observations. Dependent on the weather conditions we chose the $1\farcs0$ or $1\farcs2$ slit, resulting in spectral resolutions of we $R\approx830$ and $R\approx690$, respectively. Exposure times varied between $3\,\rm{min}$ and $15\,\rm{min}$ depending on the target magnitude as well as the atmospheric transparency.

\subsection{Data reduction}
The data were reduced using the standard long slit reduction methods within the IRAF software package \citep{Tody1986, Tody1993}. This includes bias subtraction, flat field corrections and sky subtractions using polynomial background fits along the slit direction. The last task was carried out using the \texttt{apall} routine. All observations since October 2016 were reduced using optimal extraction (weights=variance) and cosmic ray reduction within the apall routine.
Our observations resulted in low to medium signal-to-noise spectra. In all cases quasars were easily classified by their broad emission lines.
Furthermore we have used internal lamps for wavelength calibration and observed at least one spectrophotometric standard star per night. Because of changing weather conditions our absolute flux calibration may not reliable. Therefore the fluxes were scaled to match the SDSS r-band magnitudes. The spectra have not been corrected for telluric absorption features.

\section{The ELQS-S Quasar Sample}\label{sec_ELQS-S}

\begin{figure*}
 \centering
 \includegraphics[width=\textwidth]{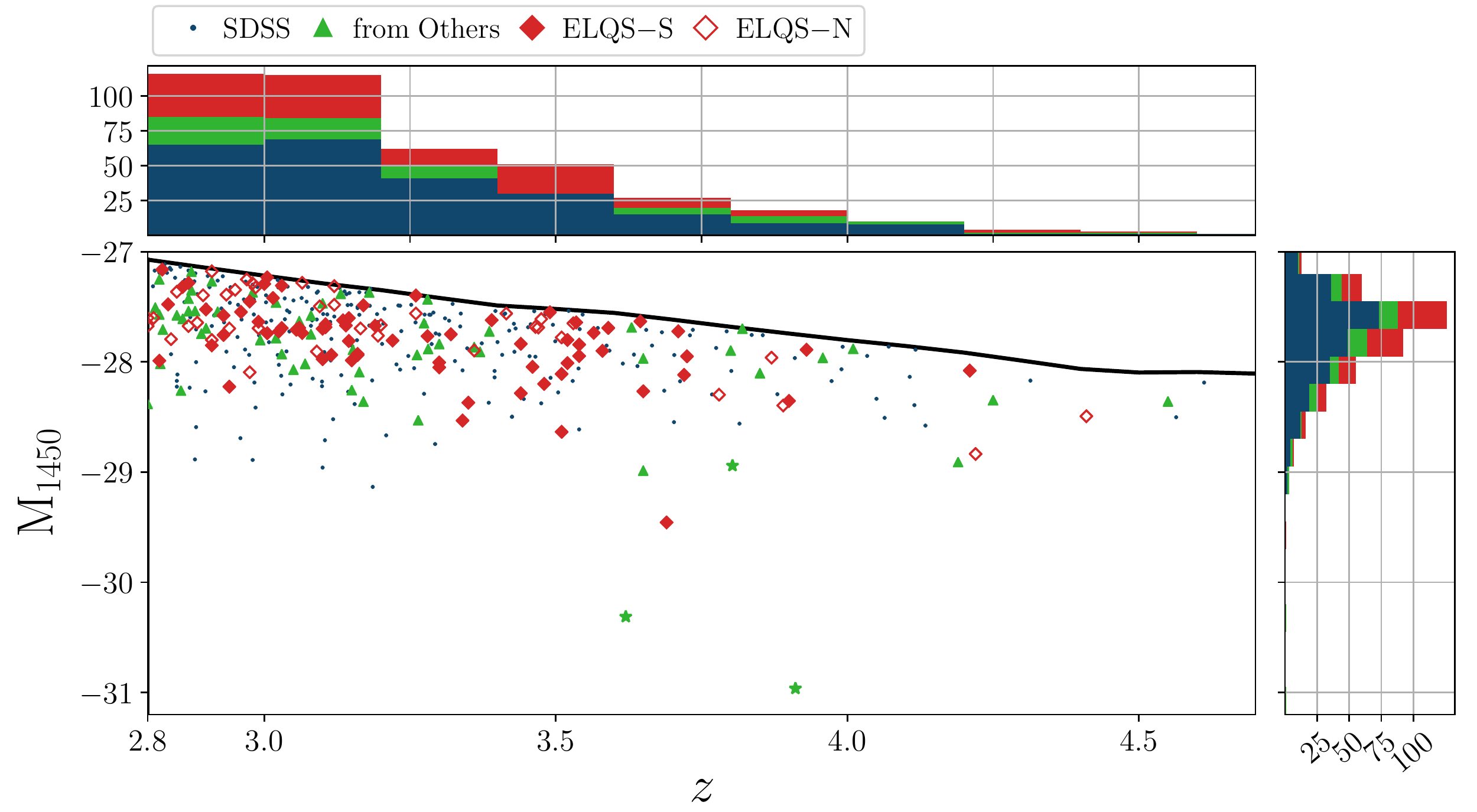}
 \caption{The distribution of all quasars in the full ELQS sample as a function of absolute $1450\text{\AA}$ magnitude ($M_{1450}$) and redshift ($z$). Quasars identified with SDSS spectroscopy or part of the SDSS DR7Q and DR14Q are shown as blue dots and labeled ''SDSS``. We also include a range of quasars that were not (re-)discovered by SDSS with identifications from the Million Quasar Catalog (MQC) or the quasar sample of Yang et al. (in preparation). These objects are depicted in green (triangles, stars). Newly identified ELQS quasars are shown as red diamonds. Solid diamonds refer to the ELQS-S sample, whereas open diamonds highlight quasars of the ELQS-N sample presented in \citetalias{Schindler2018a}.
 We also show the distribution of all quasars in histograms along both axes. The three green stars are the well known quasar lenses  Q1208+1011, B1422+231B and APM 08279+5255.}
 \label{fig_elqs_full_distribution}
\end{figure*}

\begin{table*}
\footnotesize
\centering
 \caption{ELQS primary candidate sample}
 \label{tab_cand_sample}
 \begin{tabular}{lccc}
 \tableline
 Primary candidates  & Full area & ELQS-N & ELQS-S \\
  ($m_{\rm{i}}<=18.0$ and $z_{\rm{reg}}>=2.8$)& & ($90\,\rm{deg} {<} \rm{RA} {<} 270\,\rm{deg}$) & ($\rm{RA} {>} 270\,\rm{deg}$ or $\rm{RA} {<} 90\,\rm{deg}$)\\
 \tableline
 \tableline
  Total selected primary candidates & 594 & 375 & 219 \\
 \tableline
  Good primary candidates (excluding bad photometry) & 509 & 340 & 169 \\
 \tableline
 \tableline
  Good primary candidates in the literature & 324 & 252 & 72 \\
  Good primary candidates observed & 184 & 88 & 96 \\
  Good primary candidates to observe & 1 & 0 & 1 \\
 \tableline
  Good primary candidates in the literature at $z>2.8$ & 298 & 231 & 67 \\
  Good primary candidates observed and identified as $z>2.8$ QSOs& 109 & 39 &70\\
 \tableline
 \end{tabular} 
\end{table*}

The ELQS-S sample covers the Southern Galactic Cap of the SDSS footprint ($\rm{RA} {>} 270\,\rm{deg}$ or $\rm{RA} {<} 90\,\rm{deg}$). We have selected 219 primary candidates in this area of the ELQS.
Of these, 50 candidates were discarded during visual inspection of the photometry. In most of these cases the objects were strongly blended in the WISE bands or showed photometric artifacts (bright trails identified as the source). 
Of the remaining 169 primary candidates 72 are known in the literature. These include 67 objects at $z\geq2.8$ (DR14Q:35 objects, MQC:16 objects, Yang et al.: 13, LAMOST DR2/3: 3) as well as 5 objects at $z<2.8$ (MQC:3 objects, DR14Q:1 object, SDSS spectrum: 1 object).

We obtained optical spectroscopy for 96 out of the remaining 97 unknown candidates and discovered 70 new quasars at $z\geq2.8$ and 4 new quasars with $z<2.8$. The majority of our 22 contaminants in the spectroscopic sample are stars (21), predominantly K-dwarfs (13), which have optical colors similar to the quasars in our targeted redshift range.

In total the ELQS-S catalog includes 137 quasars at $z\geq2.8$:

\begin{itemize}
 \item 70 newly identified  quasars           
 \item 35 quasars from the DR14Q
 \item 16 quasars from MQC
 \item 13 quasars from Yang et al. (2018, in preparation)
 \item 3 quasars from LAMOST DR2/3
\end{itemize}

Excluding the 50 primary candidates with unreliable photometry, we have successfully selected 137 quasars at $z\geq2.8$ out of 169 candidates. Therefore the ELQS-S sample has a selection efficiency of $\sim80\%$, consistent with the ELQS-N sample.

We show the distribution of all good primary candidates in the ELQS-S sample as a function of dereddened SDSS i-band magnitude in Figure\,\ref{fig_elqs_spec_completeness}. Known quasars from the literature are shown in blue, while new ELQS-S quasars at $z\geq2.8$ and $z<2.8$ are displayed in red and green, respectively. All objects that were spectroscopically identified not to be quasars are colored orange. In addition, two objects could not be identified. One of them was not observed and the spectrum of the other one had too low signal-to-noise to allow for a reliable classification. These two objects are shown in gray.

The figure shows a significant dip in quasar candidates around $m_{\rm{i}}\sim17.75$. It can be explained by our selection function (Figure\,\ref{fig_completeness}).
Our estimated selection completeness drops below $50\%$ at magnitudes fainter than $m_{\rm{i}}=17.5$. That the number of quasar candidates rises again in the two faintest magnitude bins is due to the intrinsic number of quasars rising at these magnitudes.
The same phenomenon is evident in our ELQS-N candidates \citepalias[see Figure\,1 in][]{Schindler2018a}. Here, there number of quasar candidates stagnates around $m_{\rm{i}}\sim17.6-17.8$ and then increases strongly beyond $m_{\rm{i}}\gtrsim17.8$.

The 70 discovery spectra of our newly identified quasars are displayed in Figure\,\ref{fig_newqso_spectra}. The spectra are ordered in redshift beginning with the lowest redshift spectrum at $z=2.82$. According to the spectroscopic redshift we highlight the positions of the broad Ly$\alpha$, \ion{Si}{4} and \ion{C}{4} emission lines with blue, orange and red bars at the top of each spectrum. The redshift as well as the designation of the object are shown in either the top right or top left corner of each spectrum. In a few cases the flux correction introduced a rising continuum at the blue end, which is likely due to insufficient signal at the bluest wavelengths. For example, J012535.83+401425.5 and J235330-050817.8 are affected by this problem.

Spectroscopic redshifts are measured by visually matching a quasar template spectrum \citep{VandenBerk2001} to the observed spectra. We estimate that the redshift uncertainty introduced by this method is $\Delta z\approx0.02$, which is accurate enough for the calculation of the QLF.

K-corrections are calculated in the same fashion as for the ELQS-N sample. We have used the sample of simulated quasar spectra \citepalias[see Section\,5.1 in][]{Schindler2018a} to derive a k-correction term as a function of redshift and magnitude to calculate the monochromatic magnitude at rest-frame $1450
\rm{\AA}$ from the SDSS i-band magnitude. 
The simulated quasar spectra were calculated on a narrow grid in redshift and absolute magnitude and k-corrections are calculated for each grid cell. This grid is then interpolated to retrieve individual k-corrections for each quasar in our sample.

Figure\,\ref{fig_elqs_full_distribution} shows the distribution of all quasars in the ELQS as a function of $M_{1450}$ and redshift. Known quasars identified by SDSS spectroscopy or included in the SDSS DR7 and DR14 quasar catalogs are shown as blue dots and labeled ''SDSS``. Other known quasars, which are part of the MQC or the quasar sample of Yang et al. (2018, in preparation) are marked with green triangles. ELQS quasars are highlighted as red diamonds, where solid diamonds refer to the new ELQS-S sample and open diamonds refer to the ELQS-N sample \citepalias{Schindler2018a}. The histograms show the binned distribution as a function of their respective axis.
The three green stars are the well known quasar lenses Q1208+1011 ($z=3.8$) \citep{Bahcall1992, Magain1992}, B1422+231B ($z=3.62$) \citep{Patnaik1992} and APM 08279+5255 ($z=3.91$) \citep{Ibata1999} and were selected as part of ELQS-N. 

For all newly discovered quasars in the ELQS-S sample we provide additional information in Table\,\ref{tab_elqs_fall_newqsos}. This includes the position in equatorial coordinates, SDSS apparent i-band magnitude, the absolute magnitude at $1450\text{\AA}$, near- and far-UV magnitudes from GALEX GR 6/7, a flag indicating visual broad absorption line quasar classification, the determined spectroscopic redshift and further notes. 

\begin{table*}
\centering
 \caption{Newly discovered quasars at $z\geq2.8$ in the ELQS-S sample}
 \label{tab_elqs_fall_newqsos}
 \begin{tabular}{ccccccccc}
  \tableline
  \tableline
 R.A.(J2000) & Decl.(J2000) & $m_{\rm{i}}$ &  $M_{1450}$ & Spectroscopic & near UV\tablenotemark{a} & far UV\tablenotemark{a} & BAL flag\tablenotemark{b} & Notes\tablenotemark{c} \\

 [hh:mm:ss.sss] & [dd:mm:ss.ss] & [mag]  & [mag] &  Redshift & [mag] & [mag] & & \\
  \tableline
00:03:39.153 & +20:31:25.80 & $17.64\pm0.01$ & -27.55 & 2.960 & - &  - & 0 & 171021 \\
00:04:28.616 & +35:20:29.04 & $17.90\pm0.01$ & -27.72 & 3.710 & - &  - & 0 & 171116 \\
00:13:11.103 & +20:53:42.74 & $17.53\pm0.02$ & -28.01 & 3.520 & - &  - & 1 & 170825 \\
00:24:48.239 & +08:12:12.03 & $16.93\pm0.01$ & -28.53 & 3.340 & - &  - & 0 & 161115 \\
00:38:11.085 & +36:40:03.95 & $17.33\pm0.02$ & -28.20 & 3.480 & - &  - & 0 & 151010 \\
00:39:01.102 & -21:44:29.23 & $17.93\pm0.02$ & -27.31 & 3.030 & - &  - & 1 & 171021 \\
00:40:21.734 & -03:34:51.36 & $17.37\pm0.02$ & -28.05 & 3.300 & $20.02\pm0.04$ &  $21.05\pm0.10$ & 0 & 161122 \\
00:52:48.631 & +21:53:25.74 & $17.87\pm0.02$ & -27.62 & 3.390 & - &  - & 1 & 170825 \\
01:00:49.245 & -03:19:13.93 & $17.24\pm0.03$ & -28.28 & 3.440 & - &  - & 0 & 161123 \\
01:04:47.159 & +25:14:22.03 & $17.39\pm0.02$ & -27.94 & 3.160 & - &  - & 0 & 161013 \\
01:08:27.246 & +28:02:18.36 & $17.81\pm0.02$ & -27.74 & 3.565 & - &  - & 0 & 171020 \\
01:15:50.139 & +26:20:15.05 & $17.31\pm0.02$ & -27.98 & 3.100 & - &  - & 0 & 161013 \\
01:17:52.184 & +05:51:24.18 & $17.73\pm0.02$ & -27.80 & 3.520 & - &  - & 0 & 171021 \\
01:18:35.607 & +39:04:58.70 & $17.92\pm0.02$ & -27.29 & 3.000 & - &  - & 0 & 171116 \\
01:20:35.946 & -09:46:32.20 & $17.57\pm0.03$ & -27.64 & 2.990 & -&  $22.18\pm0.24$ & 0 & 180123 \\
01:25:35.824 & +40:14:25.64 & $17.55\pm0.01$ & -27.71 & 3.055 & - &  - & 0 & 171020 \\
01:25:46.761 & +21:15:22.12 & $17.87\pm0.01$ & -27.89 & 3.930 & - &  - & -1 & 171021 \\
01:26:46.100 & +21:27:04.53 & $17.84\pm0.01$ & -27.28 & 2.870 & - &  - & 0 & 171021 \\
01:32:23.210 & +18:41:55.71 & $17.52\pm0.02$ & -28.12 & 3.720 & - &  - & 1 & 171010 \\
01:33:49.283 & +09:42:29.40 & $17.59\pm0.02$ & -27.58 & 2.930 & - &  - & 0 & 171021 \\
01:36:24.534 & +15:27:55.07 & $17.42\pm0.02$ & -28.00 & 3.300 & - &  - & 0 & 161013\tablenotemark{d} \\
01:38:07.139 & +17:24:14.80 & $17.66\pm0.01$ & -27.67 & 3.190 & - &  - & 1 & 171021 \\
01:48:44.807 & +25:02:02.96 & $17.41\pm0.01$ & -27.76 & 2.930 & - &  - & 0 & 170825 \\
01:55:58.279 & -19:28:49.02 & $17.33\pm0.01$ & -28.27 & 3.650 & - &  - & -1 & 161014 \\
01:59:00.676 & -03:27:37.27 & $17.62\pm0.02$ & -27.52 & 2.900 & - &  - & 0 & 171021 \\
02:02:56.078 & +31:26:20.91 & $17.57\pm0.02$ & -27.69 & 3.060 & -&  $22.42\pm0.17$ & 1 & 170825 \\
02:08:38.405 & +17:06:52.37 & $16.91\pm0.07$ & -28.63 & 3.510 & - &  - & 0 & 161123 \\
02:09:59.280 & +24:48:47.39 & $17.60\pm0.02$ & -27.68 & 3.105 & - &  - & 0 & 171021 \\
02:11:18.295 & +14:52:10.43 & $17.66\pm0.02$ & -27.90 & 3.580 & - &  - & -1 & 151011 \\
02:18:29.577 & +22:40:14.30 & $17.68\pm0.01$ & -27.62 & 3.135 & - &  - & 0 & 171020 \\
02:26:32.795 & +20:27:48.21 & $17.79\pm0.02$ & -27.32 & 2.860 & - &  - & 0 & 171109 \\
02:34:24.866 & +23:40:19.42 & $17.10\pm0.02$ & -27.99 & 2.820 & - &  - & 0 & 161122 \\
02:40:20.777 & -17:00:16.43 & $17.99\pm0.02$ & -27.40 & 3.260 & - &  - & 0 & 171007 \\
02:48:47.364 & -05:14:15.24 & $17.42\pm0.02$ & -28.11 & 3.510 & - &  - & 0 & 161014 \\
03:13:07.141 & +02:45:15.24 & $17.34\pm0.02$ & -27.99 & 3.150 & - &  - & 1 & 161014 \\
03:21:46.404 & +11:57:53.46 & $16.95\pm0.02$ & -28.23 & 2.940 & - &  - & 0 & 180120 \\
03:24:36.322 & +17:52:41.59 & $17.86\pm0.01$ & -27.69 & 3.590 & - &  - & 0 & 171021 \\
03:41:51.166 & +17:20:49.75 & $16.19\pm0.01$ & -29.46 & 3.690 & - &  - & 0 & 161218 \\
03:45:21.811 & +16:03:05.88 & $17.31\pm0.02$ & -27.85 & 2.910 & - &  - & 0 & 161122 \\
04:05:44.847 & +13:06:13.51 & $17.63\pm0.02$ & -27.48 & 2.835 & - &  - & 0 & 171021 \\
04:09:03.611 & +14:51:49.05 & $17.85\pm0.02$ & -28.08 & 4.210 & - &  - & -1 & 161123 \\
05:30:43.748 & -06:26:56.63 & $17.60\pm0.01$ & -27.95 & 3.540 & - &  - & 0 & 171006 \\
\hline
20:13:22.744 & -12:34:06.62 & $17.81\pm0.01$ & -27.42 & 3.015 & - &  - & 0 & 170518 \\
20:20:05.735 & -12:32:58.36 & $17.99\pm0.01$ & -27.23 & 3.005 & - &  - & 0 & 170518 \\
20:27:53.320 & -05:22:23.50 & $17.48\pm0.01$ & -27.74 & 3.005 & - &  - & 0 & 161014 \\
20:37:00.257 & -17:34:15.41 & $17.51\pm0.01$ & -27.81 & 3.145 & - &  - & 0 & 170518 \\
21:00:06.598 & +02:42:33.90 & $17.69\pm0.01$ & -27.95 & 3.725 & - &  - & 0 & 170518 \\
21:08:27.259 & -03:08:47.76 & $17.93\pm0.02$ & -27.16 & 2.825 & - &  - & 1 & 170518 \\
21:08:46.961 & -02:01:14.84 & $17.47\pm0.01$ & -28.04 & 3.460 & - &  - & 0 & 161014 \\
21:32:59.090 & +16:30:29.12 & $17.69\pm0.01$ & -27.84 & 3.540 & - &  - & 0 & 151011 \\
21:36:49.757 & -01:28:52.20 & $17.64\pm0.02$ & -27.77 & 3.280 & - &  - & 0 & 171021 \\
21:55:58.301 & +02:28:56.12 & $17.89\pm0.01$ & -27.64 & 3.535 & $21.58\pm0.10$ &  $23.92\pm0.48$ & 0 & 171021 \\
21:57:43.626 & +23:30:37.34 & $17.71\pm0.02$ & -27.60 & 3.145 & $20.49\pm0.16$ &  $21.39\pm0.33$ & 0 & 170825\tablenotemark{e} \\
22:15:33.089 & +23:55:54.87 & $17.52\pm0.02$ & -27.74 & 3.065 & - &  - & 0 & 171020 \\
22:16:52.892 & +30:44:51.86 & $17.54\pm0.01$ & -27.69 & 3.030 & - &  - & 0 & 161219 \\
22:32:39.479 & +13:15:18.12 & $17.97\pm0.02$ & -27.55 & 3.490 & - &  - & 0 & 171021 \\
22:32:52.174 & +34:37:12.83 & $17.95\pm0.01$ & -27.63 & 3.645 & - &  - & 0 & 171020 \\
22:38:12.040 & +33:05:46.30 & $17.51\pm0.02$ & -27.72 & 3.025 & - &  - & 0 & 171020 \\
22:42:21.472 & -03:54:58.39 & $17.40\pm0.01$ & -27.93 & 3.160 & - &  - & 0 & 171021 \\
22:46:10.792 & -06:49:53.89 & $17.84\pm0.02$ & -27.49 & 3.170 & - &  - & 1 & 171021 \\
22:55:36.224 & -02:57:36.45 & $17.56\pm0.01$ & -27.81 & 3.220 & - &  - & 0 & 171020 \\
23:02:11.049 & +03:13:44.42 & $17.68\pm0.01$ & -27.84 & 3.440 & - &  - & 0 & 171020 \\
23:13:34.613 & -10:11:52.41 & $17.75\pm0.02$ & -27.45 & 2.975 & - &  - & 1 & 171020 \\
23:22:33.545 & +17:53:09.61 & $17.69\pm0.01$ & -27.75 & 3.320 & - &  - & 0 & 170825 \\
23:24:52.615 & +18:24:16.53 & $17.10\pm0.02$ & -28.37 & 3.350 & - &  - & 0 & 171110 \\
23:31:17.250 & -05:40:21.05 & $17.64\pm0.01$ & -27.67 & 3.140 & - &  - & 1 & 171021 \\
23:33:20.540 & +12:20:22.14 & $17.59\pm0.01$ & -27.70 & 3.100 & - &  - & 0 & 170825\tablenotemark{d} \\
23:38:39.729 & +29:24:21.00 & $17.40\pm0.02$ & -28.35 & 3.900 & - &  - & -1 & 171116 \\
23:53:08.773 & +37:44:59.07 & $17.36\pm0.02$ & -27.94 & 3.115 & - &  - & 0 & 161013 \\
23:53:30.062 & -05:08:17.94 & $17.63\pm0.02$ & -27.66 & 3.105 & $20.62\pm0.20$ &  - & 0 & 171021 \\
\tableline
 \end{tabular}
\tablenotetext{1}{The near and far UV magnitudes were obtained from cross-matches within $2\farcs0$ to the GALEX GR6/7 data release}
\tablenotetext{2}{Visual qualitative BAL identification flag: $1=$BAL; $0=$no BAL; $-1=$ insufficient wavelength coverage or inconclusive archival data}
\tablenotetext{3}{This column shows the observation date (YYMMDD) and provides further information on individual objects.}
\tablenotetext{4}{These objects were also independently discovered by Yang et al.}
\tablenotetext{5}{See also HST GO Proposal 13013 (PI: Gabor Worseck), \citet{Zheng2015} and \cite{Schmidt2017}}
\end{table*}

\section{The Full ELQS Quasar Catalog}\label{sec_ELQS_QC}

The full ELQS quasar catalog is comprised of 407 objects, of which 109 are newly identified. The previously published ELQS-N catalog \citepalias{Schindler2018a}, covering only the North Galactic Cap of the SDSS footprint, included 270 (new: 39, known: 231) quasars. 
With this work we add the Southern Galactic Cap footprint of the ELQS, identifying 70 new quasars and effectively more than doubling the number of selected known quasars in this area. The selection criteria and selection function are identical to \citetalias{Schindler2018a}.
Across the entire ELQS we selected 509 primary quasar candidates of which 407 were identified to be quasars at $z\geq2.8$, resulting in an overall selection efficiency of $\sim80\%$.
The SDSS quasars make up only $\sim60\%$ of the ELQS sample and $80\%$ of all previously known quasars in SDSS footprint, if all known quasars from the literature are included.
This demonstrates that our selection is more inclusive than the SDSS quasar selection, allowing us to recover an additional 50 quasars known in the literature.

We matched the full ELQS sample against known quasar lenses. This includes a list of known quasar lenses in the NASA/IPAC Extragalactic Database (NED), the CfA-Arizona Space Telescope LEns Survey of gravitational lenses (CASTLES, C.S. Kochanek, E.E. Falco, C. Impey, J. Lehar, B. McLeod, H.-W. Rix)\footnote{\url{https://www.cfa.harvard.edu/castles/}} and the Sloan Digital Sky Survey Quasar Lens Search \citep[SLQS][]{Inada2012}. The three returned matches are the well known quasar lenses Q1208+1011 ($z=3.8$) \citep{Bahcall1992, Magain1992}, B1422+231B ($z=3.62$) \citep{Patnaik1992} and APM 08279+5255 ($z=3.91$) \citep{Ibata1999}, which were already included in the ELQS-N sample. These are highlighted as green stars in Figure\,\ref{fig_elqs_full_distribution}.

One of our candidates, J035047.55+143908.2, remains unobserved and the spectrum of another one, J025204.49+201407.9, has too low signal-to-noise to allow for an unambiguous classification. Therefore the ELQS is $99.6\%$ spectroscopically complete.

\subsection{Matches to FIRST and the Radio Loud Fraction}

We match the full ELQS sample to sources in the VLA Faint Images of the Radio Sky at Twenty-Centimeters (FIRST) catalog \citep{Becker1995} in an aperture of $3\farcs0$. We obtain measured $1.4\,\rm{GHz}$ flux densities for a total of 37 matches (ELQS-N: 34, ELQS-S: 3). All three matches in ELQS-S are to quasars already known in the literature. The full ELQS catalog (Section\,\ref{app_qso_catalog}) includes information on the match distance to the FIRST source, its $1.4\,\rm{GHz}$ peak and integrated flux density as well as the RMS error on the integrated flux density.

Since the FIRST footprint has been chosen to coincide with the SDSS North Galactic Cap footprint, we can estimate the radio loud fraction (RLF) of our ELQS-N quasar sample. This allows us to test whether the ELQS quasar sample has similar or different radio properties compared to other surveys. In \citetalias{Schindler2018a} we simply counted all sources with $1.4\,\rm{GHz}$ peak flux detections, which resulted in a RLF of $\approx12.6\%$.

\citet{Jiang2007} have analyzed the RLF for a large sample of SDSS quasars at $z{=}0{-}5$ and $-30 \leq M_{\rm{i}} < -22$. They define a radio-loud quasar based on its $R$ parameter, the ratio of the flux density at $6$\,cm ($5$\,GHz) to flux density at $2500\,\text{\AA}$ in the rest-frame,
\begin{equation}
 R = f_{6\rm{cm}} / f_{2500} \ .
\end{equation}
In their analysis they calculate $f_{6\rm{cm}}$ from the $1.4\,\rm{GHz}$ integrated flux density (if detected) by assuming a power-law slope of $\alpha={-0.5}$. They further obtain the observed flux density $f_{2500}$ at rest-frame $2500\,\text{\AA}$ by fitting a model spectrum to the SDSS broad-band photometry. Quasars are then counted as radio-loud for all values of $R\geq10$. 
They discovered that the RLF changes as a function of redshift and absolute magnitude and is well fit by
\begin{equation}
 \log_{10}\left(\frac{\rm{RLF}}{1-\rm{RLF}}\right) = b_0 + b_z (1+z) + b_M (M_{2500}+26) \ ,
\end{equation}
where $b_0 = -0.132$, $b_z=-2.052$ and $b_M = -0.183$. 

We revisit our analysis of the ELQS-N RLF by using the same criterion for radio-loud quasars as \citet{Jiang2008}. In our case we calculate the observed flux density $f_{2500}$ using the k-correction estimated from our sample of simulated quasars \citepalias[][Section\,5.1]{Schindler2018a}.  Our k-correction is not only based on a quasar continuum model, but also includes contributions from the broad quasar emission lines. The rest-frame flux density at $6$\,cm  $f_{6\rm{cm}}$ is derived identically to \citet{Jiang2007}, by assuming a power-law slope of $\alpha=-0.5$ for the k-correction.

We calculate the RLF for three different subsamples of the ELQS-N catalog restricted by $m_{\rm{i}}\leq 17.0, 17.5$ and $18.0$. In all cases we calculate median absolute magnitudes and redshifts as input into the the relation found by \citet{Jiang2007}. We compare results from the ELQS sample in Table\,\ref{tab_rlf} with the RLF calculated using the best-fit relation. 
Uncertainties on our measured RLF are derived assuming a Poisson distribution ($\sigma=\sqrt{N}$).

\begin{table}
\centering
\caption{The radio-loud fraction of the ELQS-N sample compared to the relation of \citet{Jiang2007}.}
\label{tab_rlf}
\begin{tabular}{cccc}
\tableline
\tableline
 $m_{\rm{i}}\leq$ & $17.0$ & $17.5$ & $18.0$ \\
\tableline
 $N_{\rm{ELQS}}$ & 22 & 92 & 270 \\
 $z_{\rm{median}}$ & 3.30 & 3.42 & 3.12 \\
 $M_{2500,\rm{median}}$ & -28.75 & -28.3 & -27.8 \\
 $N_{\rm{ELQS}}(R>10)$ & 2 & 10 & 25 \\
\tableline
$\rm{RLF}_{\rm{ELQS}}$ & $9.1\pm4.5\%$ & $10.9\pm2.3\%$ & $9.3\pm1.2\%$ \\ 
$\rm{RLF}_{\rm{Jiang2007}}$ & $10.6\%$ & $8.4\%$ & $8.0\%$ \\
\tableline
\end{tabular}

\end{table}

Compared to our previous estimate of $\rm{RLF}=12.6\% (m_{\rm{i}}\leq 18.0)$ in \citetalias{Schindler2018a}, where we only counted quasars with radio detections, our more rigorous RLF estimates agree with the values derived from the relation of \citet{Jiang2007} at the $1.5\sigma$ level. 
This largely confirms that the ELQS quasar sample has similar radio properties as previous SDSS surveys.

\subsection{Matches to GALEX, ROSAT 2RXS and XMMSL2}

We have cross-matched the full ELQS sample with the GALEX GR6/7 Data Release \citep{Martin2005}. Matches are evaluated within an aperture of $2\farcs0$, which corresponds to the GALEX position accuracy. For all matches we have obtained the available photometry in the near- and far-UV bands at $1350-1750\text{\AA}$ and $1750-2750\text{\AA}$, respectively. The near- and far-UV magnitudes for the ELQS-S sample are also displayed in Table\,\ref{tab_elqs_fall_newqsos}.

We obtained 55 GALEX matches to the full ELQS sample (ELQS-N: 38, ELQS-S:17). Of these matches 52 (ELQS-N: 37, ELQS-S:15) are detected in the near-UV band and 19 (ELQS-N: 10, ELQS-S:9) in the far-UV band. A subset of 16 (ELQS-N: 9, ELQS-S:7) sources were detected in both bands. 
We have discovered 109 (ELQS-N: 39, ELQS-S:70) new quasars with ELQS, of which 14 (ELQS-N: 8, ELQS-S:6) have GALEX counterparts in either or both photometric bands. 

There are three new ELQS-S quasars (J004021.734-033451.36, J215558.301+022856.12, and J215743.626+233037.34), which are detected in both near- and far-UV GALEX bands, while two objects only have far-UV photometry and one has only near-UV photometry available.

The detection of high redshift quasars in near- and far-UV bands in the observed frame suggests that their flux has not been fully absorbed by intervening neutral hydrogen along the line of sight. Thus, these objects are prime targets to study the Helium re-ionization of the universe \citep{Worseck2011, Worseck2016}.

The rate of UV detections in the full ELQS sample ($55/407\approx13.5\%$) is very similar to the rate of UV detections of our newly identified ELQS quasars ($14/109\approx13\%$). In \citetalias{Schindler2018a} we discussed the rate of UV detections in the ELQS-N sample and found that a large fraction of newly identified quasars ($8/39\approx20\%$) have UV detections compared to the overall ELQS-N sample ($38/270\approx14\%$). 

\citet{Worseck2011} found that the SDSS quasar sample preferentially selects quasars with intervening \ion{H}{1} Lyman-limit systems. 
In the case of the North Galactic Cap, where SDSS spectroscopic follow-up is completed, we would therefore expect the SDSS to have missed a larger fraction of quasars with UV detections.
Assuming our selection does not carry the same bias, we would naturally find a larger fraction of UV detections among our new ELQS-N quasars.
However, SDSS spectroscopic follow-up has not been completed in the South Galactic Cap footprint, leaving a larger and more unbiased fraction of quasars undiscovered. 
This could explain the UV detection rates of our newly identified quasars compared to the full ELQS samples.

We further cross-matched all quasars in the full ELQS catalog with pre-matched AllWISE counterparts to X-ray detections \citep{Salvato2018} from the ROSAT \citep{Truemper1982} reprocessed 2RXS catalog \citep{Boller2016} and the XMM Newton Slew 2 Survey (XMMSL2). These catalogs contain 106,573 counterparts to $0.1-2.4\,\rm{keV}$ 2RXS sources as well as 17,665 counterparts to $0.2-12\,\rm{keV}$ XMMSL2 sources. We matched the AllWISE positions of the sources in our sample to the AllWISE positions of the counterparts in a $6\farcs$ aperture.  

While we find no matches to the XMMSL2 counterparts, we recover 11 sources that have ROSAT 2RXS detections. All of them are already known quasars in the literature. The ROSAT 2RXS fluxes are included in the full ELQS quasar catalog (see Section\,\ref{app_qso_catalog}).

\subsection{Broad Absorption Line (BAL) Quasar Fraction}

We revisit our previous estimate of the fraction of broad absorption line (BAL) quasars of \citepalias[][Section\,4.2]{Schindler2018a} with the full ELQS sample. 
While a thorough quantitative analysis of the BAL quasar fraction would require the calculation of the balnicity index ($BI$) \citep{Weymann1991} or the absorption index \citep{Hall2002} from the spectral data, this is beyond the scope of this work. Traditionally BAL quasars are classified by $BI>0$. However, we limit ourself to a qualitative analysis of the BAL quasar fraction by visually classifying all ELQS quasars in BAL quasars and non-BAL quasars. Based on this classification we roughly estimate the BAL fraction of the ELQS quasar sample.

As in our previous analysis of the ELQS-N sample, we cross-match the full ELQS catalog to the SDSS DR12 quasar catalog \citep{Paris2017} and retrieve information on visual BAL quasar classifications (\texttt{BAL\_FLAG\_VI}$=1$). The DR12Q BAL flag provides information on 212 (ELQS-N: 190, ELQS-S:22) of our 407 quasars, of which 42 (ELQS-N: 40, ELQS-S: 2) are flagged as BAL quasars.

We visually inspect the spectra of all remaining objects, where available, or use previous classifications from the literature to determine their nature. 
Of all newly identified ELQS quasars 17 display BAL features. This includes 6 quasars of the ELQS-N sample and 11 new quasars of the ELQS-S ( J001311.09+205342.8, J003901.10-214429.1, J005248.64+215325.7, J013223.20+184155.6, J013807.12+172414.8, J020256.07+312620.8, J031307.14+024515.3, J210827.25-030847.8, J224610.79-064953.7, J231334.60-101152.3, J233117.24-054020.8).
A total of 7 ELQS quasars do not have sufficient signal-to-noise or wavelength coverage in their discovery spectra to allow for unambiguous classification. 
Including all quasars from the literature we could identify 79 BALs out of a sample of 384 ELQS quasars, resulting in a visual BAL quasar fraction of $\sim21\%$. For a total of 23 quasars we were not able to determine a classification, due to the lack or quality of the identification spectra.

With regard to the new ELQS-S sample, presented in this work, we have identified 22 out of 120 quasars to show BALs. Therefore the ELQS-S BAL quasar fraction is $\sim18\%$, about $4\%$ lower than in the ELQS-N sample.

The observed ELQS BAL quasar fraction of $\sim21\%$ remains high compared to previous studies in the literature \citepalias[see discussion in Section\,4.2 of][]{Schindler2018a}.
While \citet{Trump2006} find an observed traditional BAL fraction of $\sim10\%$ ($z=1.7-4.38$) in the SDSS DR3 quasar catalog, quasar samples selected from near-infrared/infrared photometry have shown to result in larger fractions of BAL quasars \citet[$\sim17.5\%$][]{Maddox2008}. 

However, it remains unclear whether our infrared based quasar selection (in the observed frame) \citep{Dai2008, Maddox2008}, our sampled redshift range, or our focus on the luminous end of the quasar distribution biases our quasar sample towards a high observed BAL fraction. 
In the future it would be interesting to conduct a more detailed analysis of the balnicity and absorption index for a large mid-infrared selected type-I quasar sample to calculate the BAL fraction as a function of redshift and absolute magnitude. Different optical quasar selection criteria applied to the mid-infrared selected quasar sample could then quantify the optical selection bias.

\section{The ELQS Quasar Luminosity Function}\label{sec_QLF}

Using the full ELQS sample, we re-evaluate our measurements of the quasar luminosity function (QLF) presented in \citetalias[][Section\,6]{Schindler2018a}. 
We calculate the binned QLF, evaluate number density and redshift evolution using a non-parametric approach and finally use a maximum likelihood method to constrain parameters for a single power-law and a broken double power-law fit to the data.
Unfortunately, we have to limit our quasar sample to the stringent photometric criteria of the 2MASS point source catalog (PSC) that we adopted for our completeness calculation \citepalias[][Section\,5.2.1]{Schindler2018a}.
We therefore have to exclude 241 quasars of our full ELQS sample, leaving 166 quasars to determine the bright-end slope of the QLF. Figure\,\ref{fig_completeness} highlights all ELQS quasars, which are included as part of the QLF sample, on top of a map of the ELQS selection function.
However, the majority of excluded quasars are at the faint end of the ELQS sample. Therefore this does not reduce the number of objects vital for the bright-end slope analysis.
Out of these 166 quasars, 38 are newly discovered and another 24 were not (re-)discovered by SDSS. Therefore this sample includes 62 quasars not part of the SDSS quasar samples, a fraction of 37.35\%.

%
%
%
%
%
%
%

\subsection{The binned QLF}\label{sec_binnedqlf}

\begin{figure*}[htp]
 \centering
 \includegraphics[width=\textwidth]{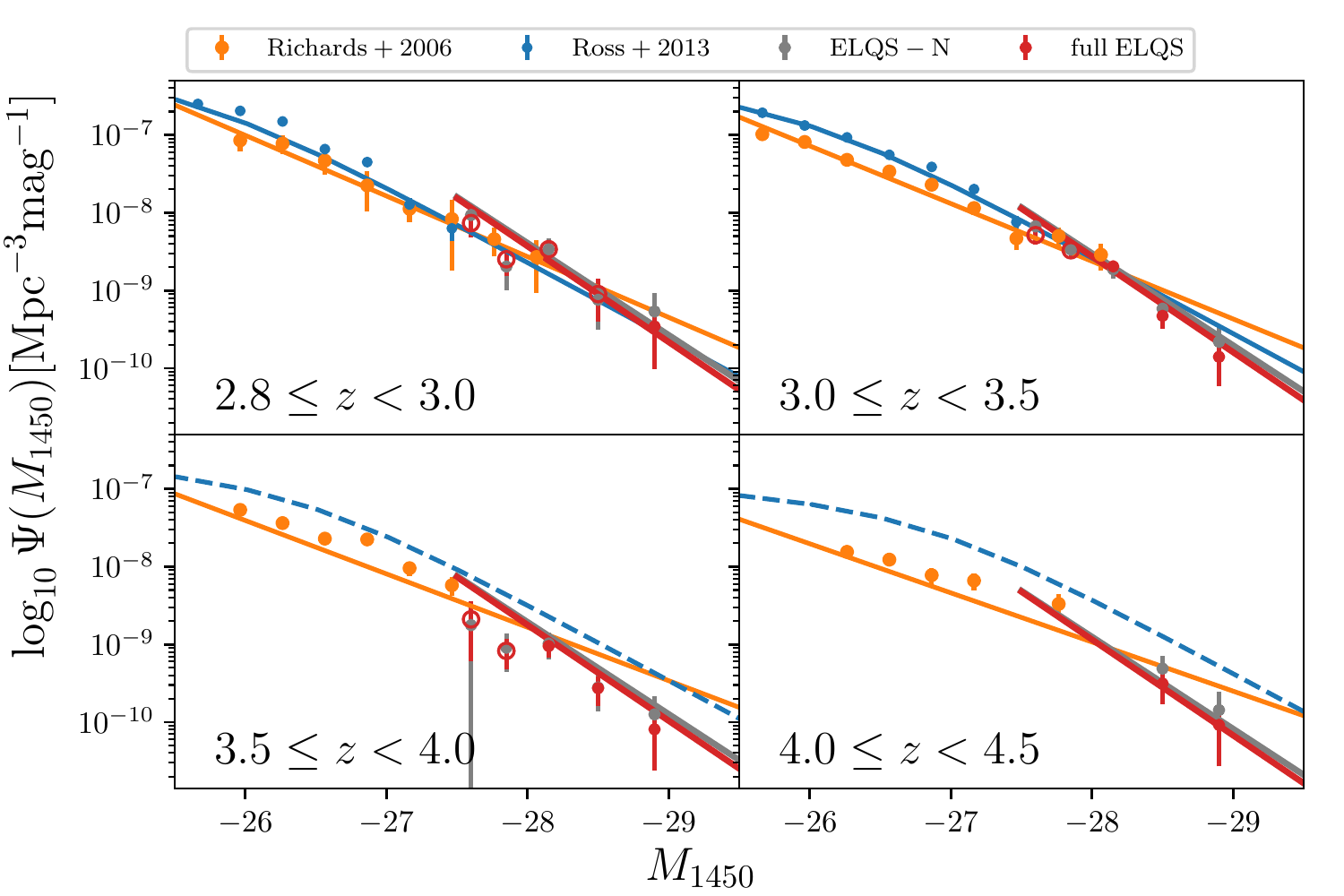}
 \caption{This figure shows the QLF of UV-bright type-I quasars as a function of absolute magnitude, $M_{1450}$, in four redshift bins. Previous results on the QLF are from the original SDSS DR3 \citep{Richards2006} (orange) and the BOSS DR9 \citep{Ross2013} (blue). The full binned ELQS is shown in red. Red open circles denote the data points that are either derived from unfilled bins or have an average completeness below $50\%$ ($N_{\rm{corr}}/N\geq2$).
 We show an earlier estimate of the QLF based on the ELQS-N sample in grey for comparison. The $1\sigma$ error bars show the purely statistical error due to the number of quasars per bin. The lines show parametric fits of the QLF to quasar distributions, where dashed lines indicate an extrapolation of the QLF prescription to higher redshifts. The red lines are from the maximum likelihood fit to the full ELQS sample (Section\,\ref{sec_mlfit}, Table\,\ref{tab_mlqlffit} first row).  The orange and blue lines correspond to the parametric fits to the QLF of \citep[][second row in their Table\,7]{Richards2006} and \citep[][PLE(first row)+LEDE(S82) in their Table\,8]{Ross2013}, respectively.}
 \label{fig_binnedqlf}
\end{figure*}

We evaluate the binned QLF over the entire ELQS footprint \citepalias[$11,838.5\pm20.1\,\rm{deg}^2$, see][]{Schindler2017} using the $1/V_{\rm{a}}$-method \citep{Schmidt1968, Avni1980} with the modification of \citet{Page2000}. 
We construct four redshift and five magnitude bins analogous to \citetalias{Schindler2018a}. The bin edges are $z=2.8, 3.0, 3.5, 4.0, 4.5$ and $M_{1450}=-29.1, -28.7, -28.3, -28, -27.7, -27.5$. We use the previously determined selection function \citepalias[][Section\,5]{Schindler2018a} to correct for incompleteness.

Figure\,\ref{fig_binnedqlf} shows the binned QLF  for the full ELQS sample (red data points, see also Table\,\ref{tab_binnedqlf}) compared to our previous estimate (ELQS-N in grey) and two other optical quasar luminosity functions determined on the SDSS DR3 \citep[orange][]{Richards2006} and DR9 \citep[blue][]{Ross2013} quasar samples. We have converted the \citet{Richards2006} and \citet{Ross2013} data, given in absolute i-band magnitudes $M_{\rm{i}}[z=2]$ continuum k-corrected to a redshift of $z=2$, to absolute magnitudes at $1450\AA$ ($M_{1450}$), assuming a spectral index of $\alpha_\nu = -0.5\ (f_\nu \propto \nu^\alpha)$.

The binned QLF of \citet{Ross2013} has been chosen to exactly match our lower two redshift bins. While the binned QLF of \citet{Richards2006} matches our three higher redshift bins, their lowest redshift bin covers $z=2.6-3.0$ compared to our coverage of $z=2.8-3.0$.

The full binned ELQS QLF is shown with filled red circles. Data points in bins that are not fully filled or where the completeness is below $50\%$ ($N_{\rm{corr}}/N\geq2$) are displayed with open red circles. These data points are prone to substantial systematic biases due to our selection function and we caution against their over-interpretation. The error bars on the binned QLF only reflect statistical uncertainties based on the detected number of quasars per bin. For comparison we show the binned QLF of the ELQS-N sample \citepalias{Schindler2018a} in grey.

Figure\,\ref{fig_binnedqlf} also displays the best fits to SDSS DR3, SDSS DR9 and ELQS quasar samples as solid lines. The color scheme follows the binned QLF. While the SDSS DR3 quasar sample \citep{Richards2006} has been fit by a single power law, the SDSS DR9 quasar sample \citet{Ross2013} extending to lower luminosities used a broken double power law parametrization. 
The ELQS sample, focused on the bright quasars, does not sample beyond the projected break of the broken double power law. Therefore, our sample can be described with a single power law. The values for the full ELQS and the ELQS-N fit are taken from Table\,\ref{tab_mlqlffit}\,(first row) and \citetalias{Schindler2018a}\,(Table\,5, first row), respectively. In all cases the fits are evaluated in the centers of the four redshift bins. We extrapolated the best fit of the SDSS DR9 QLF \citep[][see their Table\,8: PLE(first row)+LEDE(S82)]{Ross2013} beyond $z=3.5$ highlighted by the dashed line, to allow for a visual comparison in all redshift bins.

The ELQS survey allows us to extend the measurement of the QLF by one magnitude at the bright end up to $M_{1450}\approx-29$. In the brightest bin ($M_{1450}\approx-29.1$ to $-28.7$) the QLF reaches values around $10^{-10}\,\rm{Mpc}^{-3}\,\rm{mag}^{-1}$ at $z\geq3.0$. 
The data points of our binned QLF demonstrate that the bright-end slope is generally steeper as anticipated by the extrapolation of the QLF fits from \citet{Richards2006} and \citet{Ross2013} toward the brightest magnitudes. This trend is especially clear in the full ELQS sample, which results in an even steeper slope than our previous measurement based on the ELQS-N sample.

\begin{table}
\normalsize
 \centering
 \caption{The binned QLF}
 \begin{tabular}{cccccc}
  \tableline
  $M_{1450}$ & $N$ & $N_{\rm{corr}}$ & $\log_{10}\Phi$ & $\sigma\Phi$ & bin  \\
   $[\rm{mag}]$ &  & & {\footnotesize [$\rm{Mpc}^{-3}\,\rm{mag}^{-1}$] }& {\footnotesize [$\rm{Gpc}^{-3}\,\rm{mag}^{-1}$]} & filled\\
  \tableline
  \tableline
\multicolumn{6}{l}{$2.8\leq z <3.0$}\\
\tableline
-28.9 & 2 & 3.9 & -9.46 & 0.25 & True\\
-28.5 & 4 & 10.1 & -9.05 & 0.50 & True\\
-28.15 & 11 & 28.6 & -8.47 & 1.08 & True\\
-27.85 & 7 & 21.2 & -8.60 & 1.00 & True\\
-27.6 & 9 & 41.3 & -8.13 & 2.59 & True\\
\tableline
\multicolumn{6}{l}{$3.0\leq z <3.5$}\\
\tableline
-28.9 & 3 & 3.9 & -9.85 & 0.08 & True\\
-28.5 & 10 & 12.9 & -9.33 & 0.15 & True\\
-28.15 & 28 & 42.0 & -8.69 & 0.39 & True\\
-27.85 & 31 & 67.6 & -8.48 & 0.60 & True\\
-27.6 & 17 & 69.9 & -8.29 & 1.28 & False\\
\tableline
\multicolumn{6}{l}{$3.5\leq z <4.0$}\\
\tableline
-28.9 & 2 & 2.1 & -10.09 & 0.06 & True\\
-28.5 & 6 & 7.2 & -9.56 & 0.11 & True\\
-28.15 & 12 & 18.8 & -9.02 & 0.28 & True\\
-27.85 & 6 & 15.4 & -9.08 & 0.34 & False\\
-27.6 & 2 & 9.3 & -8.68 & 1.49 & False\\
\tableline
\multicolumn{6}{l}{$4.0\leq z <4.5$}\\
\tableline
-28.9 & 2 & 2.3 & -10.04 & 0.06 & True\\
-28.5 & 5 & 7.8 & -9.50 & 0.14 & True\\
\tableline
 \end{tabular}
 \label{tab_binnedqlf}
\end{table}

\subsection{The Differential Marginal Luminosity Function}\label{sec_marg_dist}

\begin{figure*}
 \includegraphics[width=\textwidth]{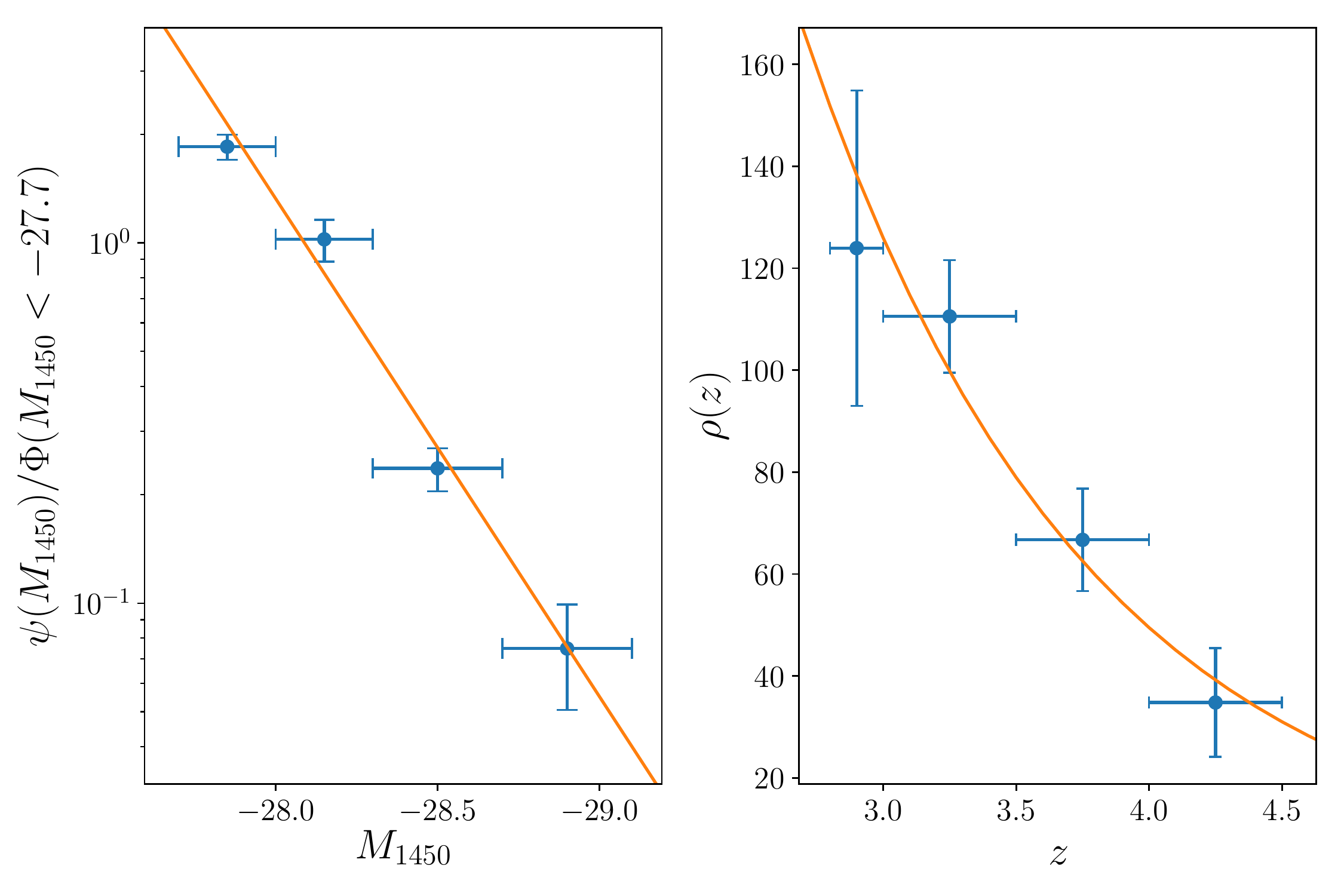}
 \caption{\textbf{Left:} The normalized marginal differential distribution of the QLF $\psi(M_{1450})/\Phi(M_{1450}\leq-27.7)$ as a function of absolute magnitude $M_{1450}$. The error bars in the magnitude direction represent the bin width, while the errors in $\psi(M_{1450})$ show the $1\sigma$ statistical error margins from the bootstrap sampling. The orange line is the maximum likelihood fit to the data points, $\psi(M_{1450})\propto 10^{-0.4(-4.45+1)\cdot M_{1450}}$. \textbf{Right:} The spatial density of the QLF $\rho(z)$, as a function of redshift $z$. The error bars in the redshift direction show the width of the redshift bins, while the error bars along $\rho(z)$ show the $1\sigma$ statistical error margins from the bootstrap sampling. The orange line is the maximum likelihood fit to the data points, $\log_{10}(\rho(z)) \propto -0.41\cdot z$.}
 \label{fig_marg_dist}
\end{figure*}

The QLF is generally a function of luminosity and redshift. Binned approaches need to divide the sample into subsamples and estimate the quasar number density per magnitude in each bin to calculate the QLF. If we can assume that the redshift and luminosity distributions in the sample are uncorrelated, we can marginalize over one variable to evaluate the marginalized QLF along the other direction, retaining a larger sample for the analysis. This is especially useful for small samples, such as ours. The assumption that the luminosity (absolute magnitude) and redshift distributions of the sample are uncorrelated is identical to assuming an underlying QLF of the form 
\begin{equation}
 \Psi(M_{1450},z) = \rho(z) \cdot \psi(M_{1450}) \ .
\end{equation}
We have introduced the methodology in the previous ELQS paper and refer all interested readers to Section\,6.2 and 6.3 of \citetalias{Schindler2018a}. 

To test whether the redshifts and luminosities of the full ELQS sample can be regarded as uncorrelated, we perform a standard correlation test \citep{Efron1992, Maloney1999, Fan2001b} and calculate the $\tau$ statistic \citepalias[][, Section\,6.2]{Schindler2018a}. As long as $|\tau|\lesssim1$, both variables can be regarded as uncorrelated parameters at the $\sim1\sigma$ level and can be treated independently. 

For the full ELQS QLF sample we obtain $\tau=-0.34$ ($\tau=-0.19$; $M_{1450}\leq-27.7$) and can therefore proceed with the calculation of the differential marginal distributions. If we restrict the sample only to higher redshifts ($3.0\leq z \leq 4.5$), $\tau$ increases to $-1.09$ ($\tau=-1.24$; $M_{1450}\leq-27.7$).

The differential marginal distributions can be calculated using Lynden-Bells $C^{-}$ estimator \citep{LyndenBell1971}. We have modified the $C^{-}$ estimator algorithm offered by the \texttt{astroML}\footnote{https://github.com/astroML} library  \citep[see also][]{Ivezic2014DataMining} to incorporate arbitrary selection functions \citep{Fan2001b}. We compute the normalized differential distributions in absolute magnitude $\psi(M_{1450})$ and redshift $\rho(z)$ with errors estimated on twenty bootstrap samples of our data. 

The marginal differential magnitude distribution $\psi(M_{1450})$, the number density of quasars as a function of magnitude, is calculated in the same magnitude bins we have chosen for the binned luminosity function in Section\,\ref{sec_binnedqlf} (starting with $M_{1450}=-27.7$). It is normalized by
\begin{equation}
\Phi(M_{1450}\leq-27.7) \equiv \int_{-\infty}^{-27.7} \psi(M_{1450})\rm{d}M_{1450}\ ,
\end{equation}
the total number of quasars with $M_{1450}\leq-27.7$.
We estimate the slope of the resulting distribution by fitting a single power law, $\log_{10}(\psi(M_{1450})) \propto - 0.4 \cdot (\beta+1) \cdot M_{1450}$, to the data.
Over the entire redshift range, $2.8\leq z \leq 4.5$, we find the slope to be best fit by $\beta=-4.45\pm0.23$.

The marginal differential redshift distribution $\rho(z)$, the spatial density of quasars as a function of redshift, uses the same redshift bins as our binned QLF analysis in Section\,\ref{sec_binnedqlf}. To analyze the evolution of the spatial density with redshift we use an exponential model, $\log_{10}(\rho(z)) \propto \gamma \cdot z$, and fit it to the data. We receive a value of $\gamma=-0.41\pm0.02$ over the entire redshift range of our sample.
The marginal differential distributions as well as their parametric fits are displayed in Figure\,\ref{fig_marg_dist}.

It should be noted that the parametric fits to both distributions combined resemble a single power law model for the QLF with exponential density evolution. We use this model in the following section to perform maximum likelihood fits.

%
%
%
%
%
%
%
%
%
%
%
%
%

\subsection{Maximum Likelihood Estimation of the QLF} \label{sec_mlfit}

\begin{table*}
\normalsize
 \centering
 \caption{Maximum Likelihood Estimation Fit Parameters for the QLF}
 \label{tab_mlqlffit}
 \begin{tabular}{ccccc}
 z & $M_{1450}^*$ & $\log_{10}[\Psi^{\star}_0]$ & $\gamma$ &  $\beta$  \\
   & [$\rm{mag}$] &  [$\rm{Mpc}^{-3}\,\rm{mag}^{-1}$]  & & \\
  \tableline
  \tableline
 2.8-4.5 & -31.5 to -27 & $-4.88^{+0.32}_{-0.32}$ & $-0.38^{+0.10}_{-0.11}$ & $-4.08^{+0.19(0.54)}_{-0.19(0.59)}$ \\
 \tableline
 3.0-4.5 & -31.5 to -27 & $-4.59^{+0.42}_{-0.41}$ & $-0.43^{+0.13}_{-0.13}$ & $-4.17^{+0.21(0.62)}_{-0.22(0.68)}$ \\
 \tableline
  2.8-4.5 & -31.5 to -28 & $-4.58^{+0.57}_{-0.56}$ & $-0.36^{+0.15}_{-0.15}$ & $-4.44^{+0.36(1.01)}_{-0.38(1.23)}$ \\
 \tableline
 3.0-4.5 & -31.5 to -28 & $-4.44^{+0.48}_{-0.47}$ & $-0.40^{+0.12}_{-0.13}$ & $-4.46^{+0.33(0.94)}_{-0.35(1.12)}$ \\
 \tableline
 \end{tabular}
\end{table*}

In this section we will calculate parametric maximum likelihood fits to the ELQS QLF sample without constraining it to redshift or magnitude bins. This analysis revisits Section\,6.4 of \citetalias{Schindler2018a} with the full ELQS sample.

We follow \citet{Marshall1983} in calculating the maximum likelihood for the QLF $\Psi(M,z)$ by minimizing the log likelihood function
\begin{equation}
\begin{split}
 S =& -2 \sum_i^N \ln\left(\Psi(M_i,z_i)(p(M_i,z_i)\right) \\
 & +2 \int \int \Psi(M,z) p(M,z) \frac{\rm{d}V}{\rm{d}z} \rm{d}M\rm{d}z \ . \label{eq_loglike}
\end{split}
\end{equation}
Confidence intervals on all parameters are derived from the likelihood function $S$ by using a $\chi^2$ distribution in $\Delta S = S-S_{\rm{min}}$\citep{Lampton1976}.

In most cases the QLF can be well represented by a double power law \citep{Boyle1988} at $z\lesssim4$, 
\begin{equation}
 \Psi(M,z) = \frac{\Psi^{\star}}{10^{0.4(\alpha+1)(M-M^{\star})}+10^{0.4(\beta+1)(M-M^{\star})}} \ . \label{eq_dplqlf}
\end{equation}
The four parameters, $\Psi^{\star}$ the overall normalization, $M^{\star}$ the break magnitude between the power laws, $\alpha$ and $\beta$ the faint and bright-end slopes, define the QLF and are known to evolve with redshift.

The break magnitude, for example, has been shown to evolve strongly from $M^{\star}_{1450}\approx-25.6$ at $z=2.8$ to $M^{\star}_{1450}\approx-26.5$ at $z=4.5$ \citep[see][their Figure\,19]{McGreer2013}. 

Therefore the ELQS sample, which probes only the luminous end of the quasar population ($M^{\star}_{1450}\lesssim-27$), does not constrain the break magnitude $M^{\star}$ nor the faint-end slope $\alpha$.

For this reason we assume a fixed break magnitude of $M^{\star}_{1450}=-26$ and parameterize the QLF using only a single power law,
\begin{equation}
 \Psi(M,z) = \Psi^{\star}(z) \cdot 10^{-0.4(\beta+1)(M-M_{1450}^{\star})} \ .
\end{equation}
We include redshift evolution by allowing the normalization $\Psi^{\star}(z)$ to vary as an exponential function of redshift,
\begin{equation}
 \log_{10}[\Psi^{\star}(z)] = \log_{10}[\Psi^{\star}_0] + \gamma \cdot z \ . \label{eq_densityevol}
\end{equation}
Here $\Psi^{\star}_0$ is the normalization at $z=0$ and $\gamma$ is a parameter of the exponential redshift evolution.

The independent redshift and magnitude evolution is supported over the full redshift range $z=2.8-4.5$ as shown by our analysis in Section\,\ref{sec_marg_dist}. 

The maximum likelihood fits are calculated using the \texttt{simqso} \citep{McGreer2013} package. We remind the reader that our ELQS QLF sample is reduced from 407 to 166 quasars by the photometric criteria we have used for the calculation of our selection function.

The parametric fits are calculated for the entire sample as well as for three subsamples constrained in redshift and/or absolute magnitude $M_{1450}$ as listed in the first two columns of Table\,\ref{tab_mlqlffit}. These ranges also serve as the integration boundaries for the calculation of $S$ in Equation\,\ref{eq_loglike}.
The remaining columns of Table\,\ref{tab_mlqlffit} list the best fit values for the three fit parameters including their $1\sigma$ statistical uncertainties. In the case of $\beta$ we have also included the $3\sigma$ uncertainties in parenthesis.
The maximum likelihood fit over the entire redshift and magnitude range (first row of Table\,\ref{tab_mlqlffit}) is also shown as the red solid line in Figure\,\ref{fig_binnedqlf}. 

For the entire sample (first row in Table\,\ref{tab_mlqlffit}), we find the bright-end slope to be steep with $\beta=-4.08$. This value is somewhat steeper than our estimate from the ELQS-N sample \citepalias[Table\,5 of][]{Schindler2018a}, $\beta=-3.96$, but lies well within the $1\sigma$ uncertainties. The single power-law fits constrain the bright-end slope at $z=2.8-4.5$ to $\beta\leq-3.4$ with a 99\% confidence.
The best fit results for the exponential density evolution, $\log_{10}[\Psi_0^\star]= -4.88$ and $\gamma=-0.38$, describe a moderately decreasing density similar to our previous estimate on the ELQS-N sample.

If we limit the ELQS QLF sample to higher redshifts (second and fourth row in Table\,\ref{tab_mlqlffit}), the bright-end slope and the density evolution steepens slightly. Imposing a faint limit of $M_{1450}=-28$ leads to a steepening of the bright-end slope, while the density evolution becomes slightly more moderate. The dependence of the bright-end slope on the sampled magnitude range potentially indicates that the break magnitude is brighter than anticipated and therefore influencing our QLF estimate. Alternatively, this effect could signal a deviation from a simple power law at the bright end.

\subsection{Double Power Law Fits}\label{sec_dplfit}

\begin{figure*}
\centering
 \includegraphics[width=\textwidth]{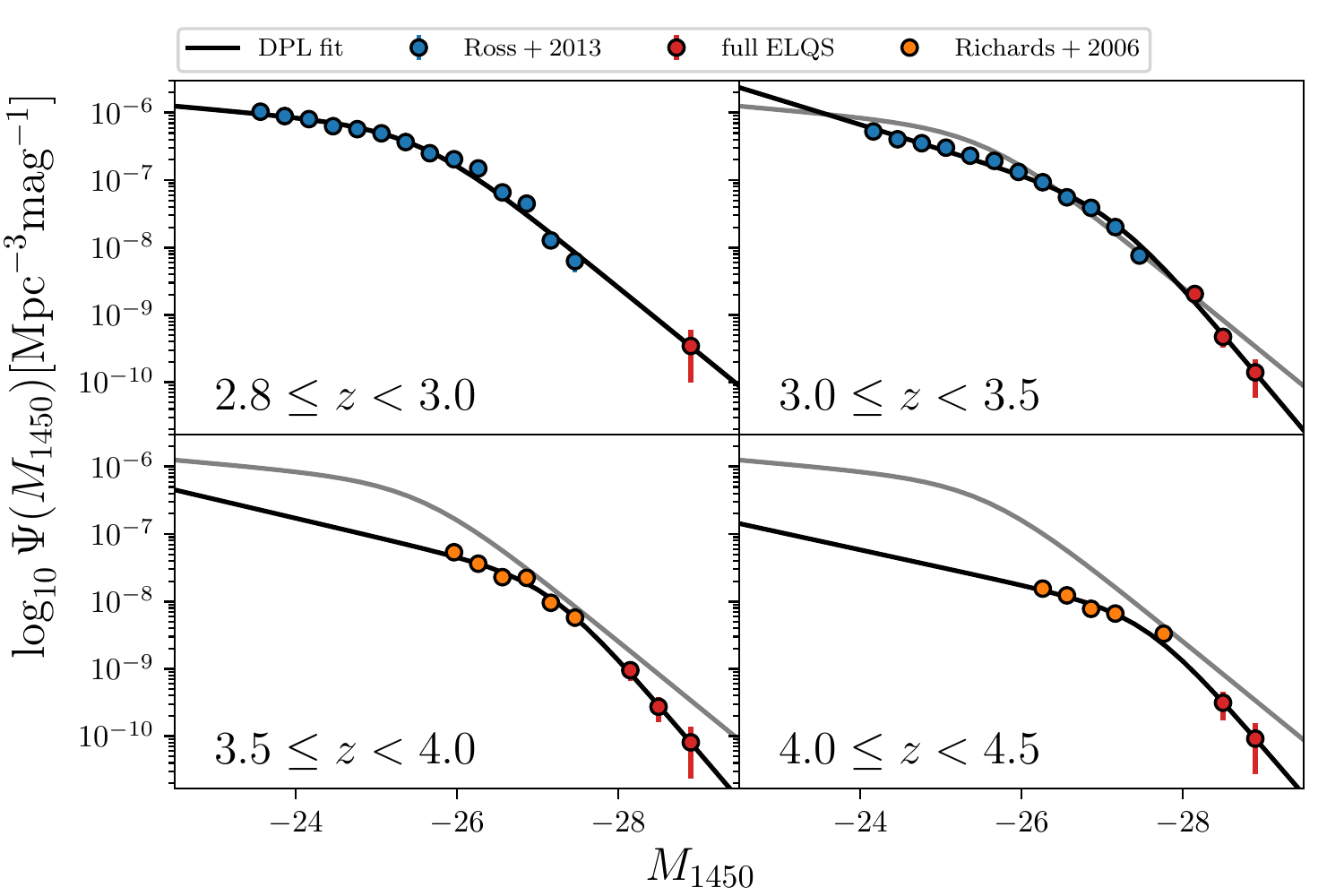}
 \caption{Broken double power-law (DPL) fits using a $\chi^2{-}\rm{minimization}$ to the binned QLFs from the ELQS (red) and the SDSS DR9 \citep{Ross2013} (blue) as well as the ELQS and the SDSS DR3 \citep{Richards2006} (orange) in the four redshift ranges sampled by the ELQS. At the lowest redshift range ($z=2.8-3.0$) the single ELQS data point does not contribute significantly to the fit. However, in the three higher redshift ranges ($z=3.0-4.5$) the ELQS data strongly help to constrain the bright-end slope. The best fit values are given in Table\,\ref{tab_dplfit_binned}. We further display the $z{=}2.8{-}3.0$ DPL fit as a grey line in the three higher redshift ranges for visual comparison.}
 
 \label{fig_qlf_dplfit}
\end{figure*}

\begin{table}
\centering
\caption{Results of maximum likelihood double power law fits to the ELQS QLF sample assuming a fixed faint-end slope ($\alpha$) and break magnitude ($M_{1450}^*$)}
\label{tab_dplfit}
 \begin{tabular}{ccccc}
 \tableline
 \tableline
 $M_{1450}^*$ & $\alpha$ & $\beta$ & $\log_{10}[\Psi^{\star}_0]$ &  $\gamma$ \\
 $[\rm{mag}]$ & & & $[\rm{Mpc}^{-3}\,\rm{mag}^{-1}]$  & \\
 \tableline
$-26.0$  & $-2.00$ & $ -4.12^{+0.18}_{-0.18}$ & $-4.88^{+0.32}_{-0.31}$ &  $-0.36^{+0.10}_{-0.10}$\\
$-26.0$  & $-1.90$ & $ -4.16^{+0.18}_{-0.18}$ & $-5.41^{+0.32}_{-0.32}$ &  $-0.38^{+0.10}_{-0.11}$\\
$-26.0$  & $-1.80$ & $ -4.15^{+0.18}_{-0.18}$ & $-5.42^{+0.32}_{-0.32}$ &  $-0.38^{+0.10}_{-0.11}$\\
$-26.0$  & $-1.70$ & $ -4.15^{+0.18}_{-0.18}$ & $-5.42^{+0.32}_{-0.32}$ &  $-0.38^{+0.10}_{-0.11}$\\

$-26.5$  & $-2.00$ & $ -4.17^{+0.18}_{-0.18}$ & $-5.40^{+0.32}_{-0.32}$ &  $-0.38^{+0.10}_{-0.11}$\\
$-26.5$  & $-1.90$ & $ -4.16^{+0.18}_{-0.18}$ & $-5.41^{+0.32}_{-0.32}$ &  $-0.38^{+0.10}_{-0.11}$\\
$-26.5$  & $-1.80$ & $ -4.15^{+0.18}_{-0.18}$ & $-5.42^{+0.32}_{-0.32}$ &  $-0.38^{+0.10}_{-0.11}$\\
$-26.5$  & $-1.70$ & $ -4.15^{+0.18}_{-0.18}$ & $-5.42^{+0.32}_{-0.32}$ &  $-0.38^{+0.10}_{-0.11}$\\
$-27.0$  & $-2.00$ & $ -4.30^{+0.18}_{-0.18}$ & $-5.95^{+0.33}_{-0.32}$ &  $-0.38^{+0.10}_{-0.11}$\\
$-27.0$  & $-1.90$ & $ -4.29^{+0.18}_{-0.19}$ & $-5.96^{+0.33}_{-0.32}$ &  $-0.38^{+0.10}_{-0.11}$\\
$-27.0$  & $-1.80$ & $ -4.28^{+0.18}_{-0.18}$ & $-5.96^{+0.33}_{-0.32}$ &  $-0.38^{+0.10}_{-0.11}$\\
$-27.0$  & $-1.70$ & $ -4.28^{+0.18}_{-0.18}$ & $-5.97^{+0.33}_{-0.32}$ &  $-0.38^{+0.10}_{-0.11}$\\
$-27.5$  & $-2.00$ & $ -4.60^{+0.21}_{-0.21}$ & $-6.46^{+0.33}_{-0.33}$ &  $-0.38^{+0.10}_{-0.11}$\\
$-27.5$  & $-1.90$ & $ -4.60^{+0.21}_{-0.21}$ & $-6.47^{+0.33}_{-0.33}$ &  $-0.38^{+0.10}_{-0.11}$\\
$-27.5$  & $-1.80$ & $ -4.60^{+0.21}_{-0.21}$ & $-6.47^{+0.33}_{-0.33}$ &  $-0.38^{+0.10}_{-0.11}$\\
$-27.5$  & $-1.70$ & $ -4.61^{+0.21}_{-0.21}$ & $-6.47^{+0.33}_{-0.33}$ &  $-0.38^{+0.10}_{-0.11}$\\
$-28.0$  & $-2.00$ & $ -5.24^{+0.30}_{-0.31}$ & $-6.88^{+0.34}_{-0.33}$ &  $-0.40^{+0.10}_{-0.10}$\\
$-28.0$  & $-1.90$ & $ -5.26^{+0.30}_{-0.31}$ & $-6.87^{+0.34}_{-0.33}$ &  $-0.41^{+0.10}_{-0.10}$\\
$-28.0$  & $-1.80$ & $ -5.29^{+0.30}_{-0.31}$ & $-6.86^{+0.34}_{-0.33}$ &  $-0.41^{+0.10}_{-0.10}$\\
$-28.0$  & $-1.70$ & $ -5.32^{+0.30}_{-0.31}$ & $-6.85^{+0.34}_{-0.33}$ &  $-0.41^{+0.10}_{-0.10}$\\
\tableline
 \end{tabular}
\end{table}

\begin{table*}
 \centering 
\caption{Result of double power law fits ($\chi^2{-}\rm{minimization}$) to the binned QLFs from ELQS/SDSS DR9 \citep{Ross2013} and ELQS/SDSS DR3 \citep{Richards2006} (see Figure\,\ref{fig_qlf_dplfit}).} 
\label{tab_dplfit_binned}
\begin{tabular}{ccccccc}
\tableline
\tableline
 z & $M_{1450}^*$ & $\alpha$ & $\beta$ & $\log_{10}[\Psi^{\star}_0]$ & $\sigma[\Psi^{\star}_0]$& ELQS Combined With\\
   & $[\rm{mag}]$                    &  & & $[\rm{Mpc}^{-3}\,\rm{mag}^{-1}]$ &  $[\rm{Gpc}^{-3}\,\rm{mag}^{-1}]$ & \\
\tableline

 2.8-3.0 & $-25.58\pm0.22$ & $-1.27\pm0.20$ & $-3.44\pm0.07$ & $-6.23$ & 185.93 & \citet{Ross2013}\\
 3.0-3.5 & $-27.13\pm0.21$ & $-1.92\pm0.16$ & $-4.58\pm0.18$ & $-7.33$ &  22.07& \citet{Ross2013}\\
 3.5-4.0 & $-27.17\pm0.28$ & $-1.70\pm0.66$ & $-4.52\pm0.15$ & $-7.65$ & 15.51& \citet{Richards2006}\\
 4.0-4.5 & $-27.57\pm0.24$ & $-1.65\pm0.46$ & $-4.50\pm0.18$ & $-8.16$ & 3.96& \citet{Richards2006}\\
\tableline
\end{tabular}
\end{table*}

To ascertain the influence of the break magnitude on our single power law fits, we investigate how our data would be represented assuming a broken double power law (DPL) for the QLF \citep{Boyle1988, Pei1995}. 


At first we perform QLF fits following the broken DPL (Equation\,\ref{eq_dplqlf}) with density evolution (Equation\,\ref{eq_densityevol}) assuming a large range of fixed values for the break magnitude, $M_{1450}^*= -26.5, -27, -27.5, -28$, and the faint-end slope, $\alpha= -1.7, -1.8, -1.9, -2.0$. The choices for the fixed parameters are guided by previous works at lower and higher redshifts \citep{Croom2009, Ross2013, McGreer2013, Yang2016}. The best fit parameters of the resulting 16 fits are listed in Table\,\ref{tab_dplfit}.


While the assumed faint-end slope does not have any strong effect on the three fitted parameters ($\beta, \log_{10}[\Psi^{\star}_0],  \gamma$), the break magnitude clearly does affect the bright-end slope and the normalization.
For brighter assumed break magnitudes we obtain a lower normalization and a steeper bright-end slope, revealing the potential bias our single power-law fits carry. The redshift evolution of the normalization, $\gamma$, is not affected by different assumptions of the break magnitude.
The dependence of the bright-end slope and the normalization on the break magnitude is already well documented in the literature \citep{McGreer2013, Yang2016} and is a degeneracy that arises from the functional form of the broken double power law. 
That our fit results are not affected by the choice of the faint-end slope only reflects that our data does not constrain the faint-end slope. Other studies of the QLF that constrain the faint-end slope find a dependence on the break magnitude \citep{McGreer2013, Onoue2017}.

In addition to the sixteen double power law fits with fixed break magnitude and faint-end slope, we calculate a fit to a QLF model with additional evolution in the break magnitude (luminosity evolution),
\begin{equation}
 M_{1450}^*(z) = M_{1450}^*(z=2.9) + c \cdot (z-2.9) \ . \label{eq_lumevol}
\end{equation}
This redshift parametrization of the QLF with separate luminosity evolution and density evolution is often abbreviated as LEDE. In our case we assume $M_{1450}^*(z=2.9)=-27.0$, $c=-0.3$ and $\alpha=-1.9$. The choice of $c$ was motivated by the binned QLF of \citet{Richards2006} and \citet{Ross2013} at the faint end. We retrieve best fit values of  $\beta=-4.33^{+0.18}_{-0.19}$ $\log_{10}[\Psi^{\star}_0] = -4.86^{+0.32}_{-0.32}$, and $\gamma=-0.75^{+0.10}_{-0.10}$. 
The value for the bright-end slope and the normalization are still consistent with our single power law results, while the density evolution parameter $\gamma$ has steepened significantly. However, with the assumption of a relatively flat slope for the luminosity evolution, our results for the density evolution are similar to \citet{Yang2016}, who fit a LEDE model at $z\sim5$ to find $c=-0.5\pm0.08$ ($c_2$ in their notation) and $\gamma=-0.81\pm0.03$ ($c_1$ in their notation).


Additionally, we perform broken DPL fits to the binned QLF of our sample complemented by the QLFs of \citet{Richards2006} at $z=3.5-4.5$ and \citet{Ross2013} at $z=2.8-3.5$. We show the best-fit results in Figure\,\ref{fig_qlf_dplfit} and present the best-fit values in Table\,\ref{tab_dplfit_binned}. 
The fit in the lowest redshift range, $z=2.8-3.0$, is dominated by the \citet{Ross2013} QLF data as we contribute only one data point at the bright end. At the higher redshifts the binned ELQS QLF contributes substantially to the determination of the bright-end slope and the break magnitude as it extends the dynamic range in $M_{1450}$ by more than one magnitude towards the bright end.

At $z=3.0-4.5$ the best-fit break magnitudes have values of $M_{1450}^*\leq-27$ and are therefore not only in the magnitude range sampled by the ELQS, but also around one magnitude brighter than expected from the literature \citep{McGreer2013}. It is not surprising that the bright-end slopes of the broken DPL fits are subsequently steeper than the single power-law fits. At $z=3.0-3.5$, $z=3.5-4.0$, and $z=4.0-4.5$, we retain bright-end slopes of $\beta= -4.58, -4.52,$ and $-4.50$ for the broken DPL, which are significantly steeper than our best-fit single power-law slope of $\beta\approx-4.1$. 

These results strongly indicate that the ELQS extends to the break magnitude at $z=2.8-4.5$. Complementing the ELQS data with previous measurements of the binned QLF leads to the conclusion that the bright-end slope is even steeper than our single power-law fits suggested.

\section{Discussion}\label{sec_discussion}

\begin{figure*}
 \includegraphics[width=\textwidth]{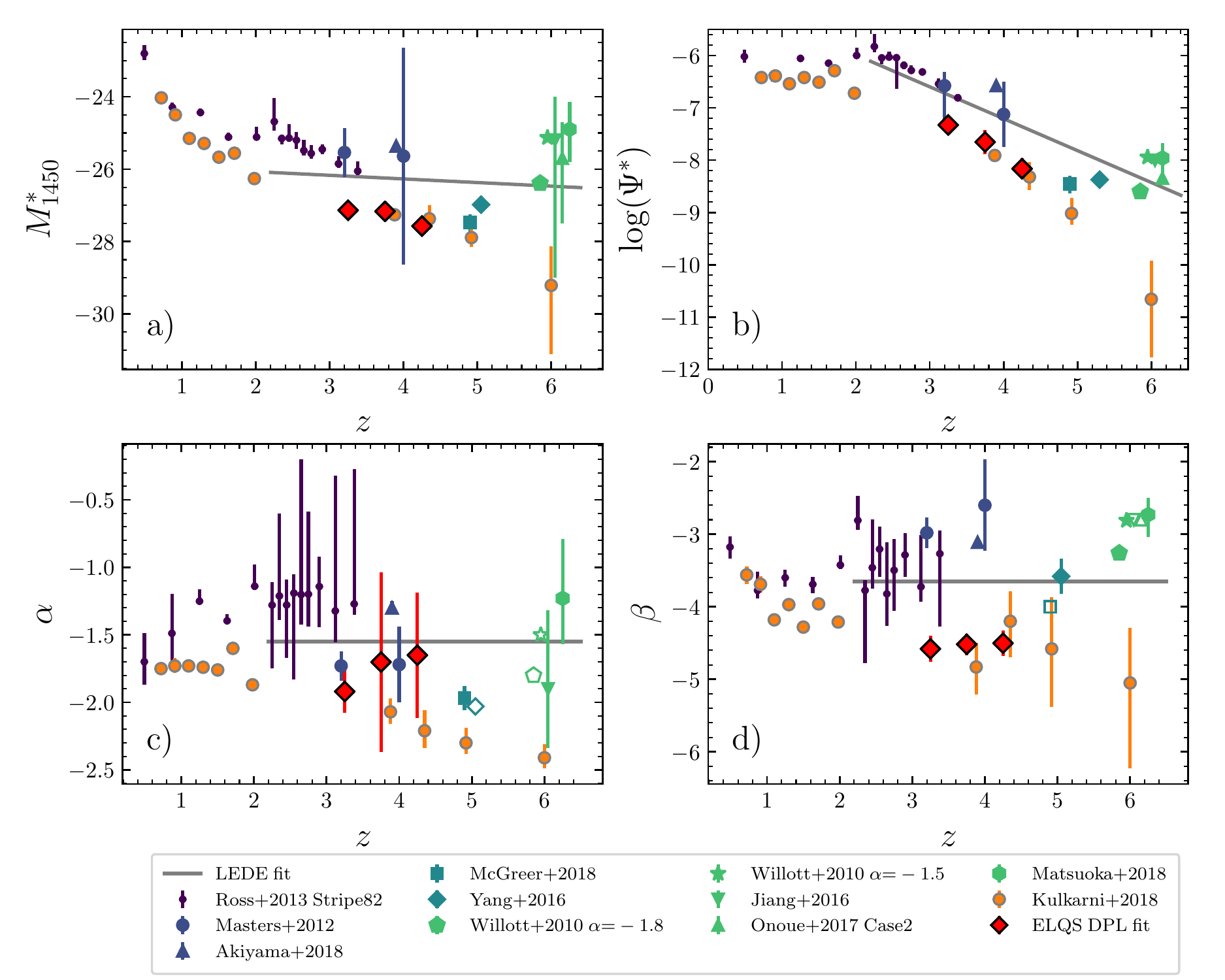}
 \caption{QLF parameters for fits of a broken double power law in narrow redshift bins. Solid data points are results from fits to data, while open data points symbolize fixed values in the QLF fit. We compare our double power-law fit results (red diamonds, see Figure\,\ref{fig_qlf_dplfit} and Table\,\ref{tab_dplfit_binned}) with a variety of other studies across the whole redshift range \citep{Ross2013, Masters2012, McGreer2018, Yang2016, Willott2010a, Jiang2016, Onoue2017, Kulkarni2018, Akiyama2018, Matsuoka2018_arxiv}. In addition, the LEDE best fit results (Section\,\ref{sec_lede_fit}) are displayed as the grey solid line.
 (a) The break magnitude, $M_{1450}^*$, which brightens with increasing redshift. (b) The density normalization, $\log_{10}[\Psi^{\star}]$, which decreases strongly with increasing redshift. 
 (c) The faint-end slope, $\beta$, which shows no systematic trend with redshift. (d) The bright-end slope, which also does not show any consistent redshift-dependent behavior.
}
 \label{fig_qlf_param_comp}
\end{figure*}

Our analysis of the ELQS sample encourages a significantly steeper bright-end slope ($\alpha\approx-4.1$ to $-4.7$) than previous studies at these redshifts suggested \citep[$\alpha\approx-2.5$;][]{Richards2006, Fan2001b, Masters2012, Akiyama2018}. In this section we will place our results in context with other studies of the QLF across the whole redshift range probed.

In most cases a broken double power law form is fit to the QLF of UV-bright type-I quasars in narrow redshift slices to determine the four fit parameters ($M_{1450}^*, \alpha, \beta, \log_{10}[\Psi^{\star}]$) as a function of redshift. While the ELQS does not probe the faint end of the QLF, we have combined our binned ELQS QLF with previous measurements from the SDSS \citep{Richards2006, Ross2013} to calculate all four fit parameters. 

We exclude the redshift range $z=2.8-3.0$ from the following comparison, because the ELQS only contributes one data point with reasonably large uncertainties to the fit. Therefore the best fit results only reflect the \citet{Ross2013} data.

Figure\,\ref{fig_qlf_param_comp} shows our best fit results (red diamonds) compared to a variety of other studies. Open data points illustrate values that were held fixed in the fitting process. 
At the lowest redshifts we display the data of Stripe 82 in \citet{Ross2013} (purple dots). At intermediate redshifts we compare our data with the study of \citet{Masters2012} (blue circles) and \citet{Akiyama2018} (blue triangles). The former study is focused on a faint sample of known quasars and photometric quasar candidates from the Cosmic Evolution Survey (COSMOS), which has to rely on the SDSS DR3 quasar sample of \citet{Richards2006} at the bright end. Similarly, \citet{Akiyama2018} probe the faint end of the $z\approx4$ QLF using the Hyper Surprime-Cam Wide Survey and use data from the SDSS DR7 to extend their sample to the bright end.
At $z\sim 5$ we compare to data from \citet{Yang2016} and \citet{McGreer2018} (turquoise squares and diamonds, respectively). The former study constrains the faint-end slope, while the latter analyses the bright end. 
All light green data is from studies at $z\sim6$ \citep{Willott2010a, Jiang2016, Onoue2017, Matsuoka2018_arxiv}. The data is slightly offset in redshift for display purposes. For some of the parameters at the highest redshifts uncertainties are not given in the corresponding publication.
\citet{Kulkarni2018} have recently reanalyzed a multitude of quasar samples across $z=0-7.5$ to determine the evolution of the QLF in a homogeneous way. For their analysis of the QLF evolution they exclude data at the lowest redshifts ($z\leq0.5$) and around $z=2.4-3.8$, where they argue that the quasar samples are significantly affected by observational biases. We display their DPL fit parameters a orange circles at the redshifts unaffected by these systematics (see filled symbols in their Figure\,4).

Figure\,\ref{fig_qlf_param_comp} (a) displays the redshift evolution of the break magnitude. Following the data of \citet{Ross2013}, \citet{Yang2016} and \citet{McGreer2018} one can make out a clear trend of the break magnitude decreasing with increasing redshift. The data of \citet{Masters2012} and the studies at the highest redshift have significant error bars (where available), possibly allowing for the general trend to be continued up to $z\sim6$. 
Our best-fit break magnitudes are clearly offset from the general trend by about one magnitude towards the brighter end. However, they are in good agreement with the data of \citet{Kulkarni2018}.

The normalization, $\log_{10}(\Psi^*)$, shown in Figure\,\ref{fig_qlf_param_comp} (b), decreases strongly with increasing redshift. The  results of \citet{Kulkarni2018} generally follow the same trend as the other literature values until $z\sim5$, while lying lower at all redshifts. They differ substantially from the results at $z\sim6$, which is a consequence of the brighter break magnitude as we discuss later.
Our best-fit results lie below the general trend and therefore agree well with the data of \citet{Kulkarni2018}. The best-fit value of the normalization is strongly dependent on the break magnitude. The agreement of our data with the results of \citet{Kulkarni2018} is therefore not surprising.

Figure\,\ref{fig_qlf_param_comp} (c) shows the faint-end slope of the different studies as a function of redshift. The data do not suggest a strong evolution of the faint-end slope with redshift. While the purple data \citep{Ross2013} suggests that the faint-end slope seems to be flattening with redshift, this trend is not supported by the other literature. The data of \citet{Kulkarni2018} even show the faint-end slope to be steepening with redshift. We retain faint-end slopes around $\alpha\approx-1.8$, which show considerable uncertainties. However, these values are in general agreement with the other data.

In Figure\,\ref{fig_qlf_param_comp} (d) we compare the best-fit bright-end slope of our work with the other values in the literature. Similarly to the faint-end slope there is no evident evolution of the bright-end slope with redshift. 
Some earlier studies reported a flattening of the bright end slope from $z=0$ towards $z\approx4$ \citep{Richards2006, Fan2001b}. The data of \citet{Masters2012} and \citet{Akiyama2018}, seem to support this claim, while our results and the \citet{Kulkarni2018} data do not support it and rather argue for a consistently steep bright-end slope.


Figure\,\ref{fig_qlf_dplfit} underlines the importance of the ELQS sample on the determination of the QLF compared to the original \citet{Richards2006} and \citet{Ross2013} samples.
Because we extended the bright end by one magnitude, we are now able to securely constrain the bright-end slope and the break magnitude. The break magnitude is about one magnitude brighter than previously expected \citep{McGreer2013, Yang2016}, which has a strong impact on the measured faint- and bright-end slopes. As a result we find best-fit bright-end slopes around $\beta\approx-4.6$ over $z=3.0-4.5$.

Our results generally agree well with the recent re-estimation of the QLF of \citet{Kulkarni2018}. In their work the authors combine a large range of quasar samples to study the QLF evolution from $z=0$ up to $z=7.5$. In the redshift range probed by the ELQS they rely on the SDSS DR7 quasar sample \citep{Schneider2010}, the SDSS DR9 quasar sample \citep{Ross2013} and the \citet{Glikman2011} quasar sample.
The ELQS quasar sample overlaps strongly with the SDSS DR7 and DR9 quasars samples as we cover the same footprint. However, our novel quasar selection is independent from the SDSS quasar selection methodology. Therefore, the ELQS QLF analysis can be considered an independent measurement with regard to the work of \citet{Kulkarni2018}.

If we were to assume the evolution of the break magnitude evident in the study of \citet{Kulkarni2018} and our work, the break magnitude would reach values of $M_{1450}^*\approx-29$ at $z=6$, making constraints on the bright-end slope above $z\approx6$ increasingly inaccessible. 
However, how does one reconcile this result and the studies at the highest redshifts \citep{Matsuoka2018_arxiv, Wang2018_arxiv} that find the QLF well represented by a DPL with a much lower break magnitude around $M_{1450}^*\approx-25$ and a flatter bright-end slope around $\beta\approx-2.6$? Is their bright-end slope the faint-end slope at lower redshifts and thus we need to introduce a broken triple power law for the QLF? Or is there an entirely different functional form that can describe the QLF better across all redshifts?
While the recent results allowed us to raise these questions, an obvious solution is not yet in sight.

\subsection{Comparison to an evolutionary QLF model fit}\label{sec_lede_fit}

We perform an evolutionary fit to the binned QLF at higher redshifts ($z>2.2$) to compare with the DPL fits of the ELQS at $z=2.8-4.5$. For this analysis we supplement  the binned QLF data used above for the DPL fits with the most recent data at lower \citep{Palanque2016} and higher \citep{Akiyama2018, McGreer2018, Matsuoka2018_arxiv} redshift.

To describe the evolution of the QLF across larger redshift ranges it is common practice to introduce redshift dependencies on the parameters of the double power law form.  We adopt an independent luminosity and density evolution model \citep[LEDE, see also][]{Ross2013} to describe the redshift dependence of the DPL. The LEDE model has been successful in describing the evolution of the QLF at higher redshifts \citep{Ross2013, Palanque2016}.

We adopt the  parametrization of \citet{Ross2013} and model the evolution of the normalization and break magnitude as
\begin{equation}
\log_{10}[\Psi_0^*](z) = \log_{10}[\Psi_0^*](z=2.2) + c_1 \cdot (z - 2.2) \ ,
\end{equation}
and 
\begin{equation}
M_{1450}^{*}(z) = M_{1450}^{*}(z=2.2) + c_2 \cdot (z - 2.2) \ .
\end{equation}
The QLF is then fully described by the normalization at $z=2.2$, $ \log_{10}[\Psi_0^*](z=2.2)$, the break magnitude at $z=2.2$, $M_{1450}^{*}(z=2.2) $, the power law slopes $\alpha$ and $\beta$, and the two evolutionary parameters, $c_1$ and $c_2$.

We perform maximum likelihood fits to the binned QLF data using \texttt{emcee} \citep{emcee} for Markov Chain Monte Carlo sampling of the parameter space. We adopt the median values of the posterior distributions as our best-fit values and summarize them in Table\,\ref{tab_lede_fit}.

We show the LEDE best-fit model (solid black line) in Figure\,\ref{fig_lede_fit} compared to the binned QLF data, our DPL fits in individual redshift ranges (solid red line), and the best-fit DPLs of the high-redshift QLFs (solid colored lines) according to their respective studies.
While the LEDE fit seems to be an adequate representation of the binned QLF data, it deviates from the double power law fits in the specific redshift bins. 
If we compare the LEDE fit to the fit of \citet{McGreer2018} and our DPL fits (red solid lines, see Section\,\ref{sec_dplfit}), the LEDE model's break magnitude is fainter and it's bright end slope is less steep. This situation is reversed, when comparing the LEDE model to \citet{Akiyama2018} and \citet{Matsuoka2018_arxiv}.

This comparison highlights the disparities between the different studies at different redshift ranges, which already became clear in Figure\,\ref{fig_qlf_param_comp}. In light of these differences in break magnitudes and bright- and faint-end slopes, it becomes increasingly challenging to find one model that coherently describes the redshift evolution of the QLF.

\begin{figure*}
\includegraphics[width=0.9\textwidth]{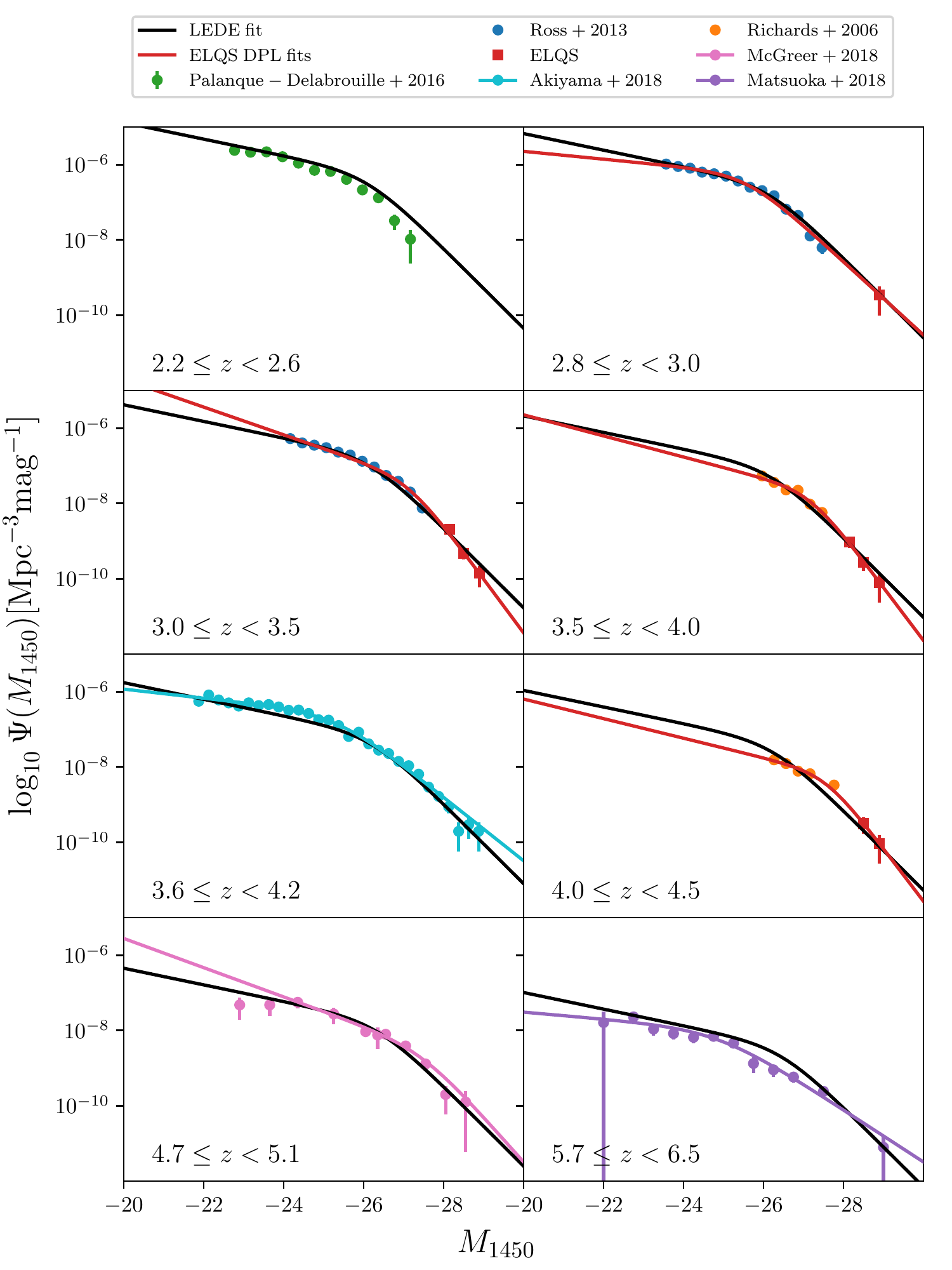}
\caption{Results of the LEDE fit (solid black line) to binned QLF data (colored data points) across $z=2.2-6.5$. The binned QLF data is taken from a variety of recent studies \citep{Richards2006, Ross2013, Palanque2016, Akiyama2018, McGreer2018, Matsuoka2018_arxiv}.We further display the DPL fit to the binned QLF in individual redshift bins as solid lines, colored according to their reference.}
\label{fig_lede_fit}
\end{figure*}

\begin{table}
\centering
\caption{LEDE model fit parameters}
\begin{tabular}{cc}
\tableline 
Best fit parameter & Best-fit value \\
\tableline 
\tableline
 $ \log_{10}[\Psi_0^*](z=2.2)$ & $-6.11\pm 0.03$ \\
 $M_{1450}^{*}(z=2.2)$ &  $-26.09\pm0.05$\\
 $\alpha$ & $-1.55\pm0.02$\\
 $\beta$ & $-3.65\pm0.06$\\
 $c_1$ &  $-0.61\pm 0.02$\\
 $c_2$ & $-0.1\pm-0.03$\\
 \tableline
\end{tabular}
\label{tab_lede_fit}
\end{table}

\section{Conclusions}\label{sec_conclusion}

We have presented the motivation for the Extremely Luminous Quasar Survey (ELQS) as well as our novel quasar selection using a JKW2 color cut and machine learning methods (random forests) in \citetalias{Schindler2017}. A subsequent publication, \citetalias{Schindler2018a}, reported our first spectroscopic observations in the North Galactic Cap (ELQS-N), constrained the ELQS selection function, and discussed a preliminary analysis of the QLF based on the ELQS-N quasar sample. With this work we conclude the ELQS. Spectroscopic follow-up of ELQS candidates has been mostly completed, allowing us to present the full ELQS quasar catalog (Section\,\ref{sec_ELQS_QC}) and analyze the QLF on the full ELQS QLF sample: 

\begin{enumerate}
 \item We report the discovery of 70 new quasars (see Table\,\ref{tab_elqs_fall_newqsos} and Figure\,\ref{fig_newqso_spectra}) at $z=2.8-4.5$ as part of the ELQS Southern Galactic Cap sample (ELQS-S). The full ELQS-S sample contains 137 quasars over an area of $4,237.3\pm12.9\,\rm{deg}^2$ of the SDSS footprint at $\rm{RA} {>} 270\,\rm{deg}$ and $\rm{RA} {<} 90\,\rm{deg}$ . Our newly discovered quasars double the known population of quasars in the South Galactic Cap footprint of the SDSS survey. This sample improves upon the known SDSS spectroscopic incompleteness of the South Galactic Cap allowing for a more unbiased measurement of quasars across the full SDSS footprint.
 
 \item The full ELQS quasar catalog is comprised of 407 quasars of which 109, or $\sim26\%$, are newly identified. Only 239 of the already known 298 quasars are part of SDSS DR14Q, which means that our quasar selection includes an additional 59 quasars from the literature missed by the SDSS quasar surveys. Overall our selection identified 509 primary quasar candidates, of which 407 were identified to be quasars, resulting in a selection efficiency of $\sim80\%$.
 
 \item We have cross-matched the full ELQS sample to the AllWISE counterparts of the reprocessed ROSAT 2RXS catalog, GALEX GR6/7 and the FIRST survey. There are 11 sources with ROSAT 2RXS detections, which are all already known quasars in the literature. We identified 55 matches to GALEX sources, of which 14 are newly identified quasars with GALEX counterparts in either or both photometric UV bands. The rate of UV detections is about $13\%$ for the full sample. We were also able to obtain 1.4\,GHz flux measurements from FIRST for 37 quasars in the full ELQS sample. Since the FIRST footprint has been chosen to overlap with the SDSS North Galactic Cap footprint, we evaluated the radio loud fraction (RLF) for the ELQS-N sample. We use the $R$ parameter, the rest frame ratio of the flux density at $6$\,cm ($5$\,GHz) to flux density at $2500$\,\AA, to classify quasars as radio loud with $R\geq10$. We estimate an $\rm{RLF}(m_{\rm{i}}\leq18) \approx 9.3\pm1.2\%$ for the ELQS-N sample, which generally agrees with the value derived from the relation of \citet{Jiang2007} for our median redshift and absolute magnitude ($\rm{RLF}_{\rm{Jiang2007}}=8.0\%$).
 
 \item We further determine the fraction of quasars with broad absorption lines (BALs) for the full ELQS sample. The DR12Q BAL flag (BAL\_VI) provides information for 212 quasars and we classify the remaining quasars by visual inspection or use previous classifications in the literature. A total of 23 quasars could not be identified due to the lack or the quality of the identification spectra. We could identify 79 BAL quasars out of 384 sources in the full ELQS sample. Of all newly identified quasars 17 display BAL features. We estimate an observed BAL quasar fraction of $\sim21\%$, which is large compared to other quasar samples \citep[for example][]{Trump2006, Maddox2008}. It remains unclear if the larger BAL quasar fraction is due to our observed-frame near-infrared based selection, the sampled redshift range \citep[redshift dependence,][]{Allen2011} of the ELQS or our focus on the luminous end of the quasar distribution (luminosity dependence). 
 
 \item We evaluate the QLF based on the full ELQS sample in Section\,\ref{sec_QLF}. A comparison of our binned QLF to the SDSS DR3 \citep{Richards2006} and SDSS DR12 \citep{Ross2013} QLF (see Figure\,\ref{fig_binnedqlf}) shows that the bright-end slope is steeper than the parametric fits to these two references suggest. We continue to analyze the differential marginal distributions of the QLF along the luminosity and redshift variable (Figure\,\ref{fig_marg_dist}), finding a steep bright-end slope of $\beta\approx-4.45$ for a single power law parametrization. A maximum likelihood fit to a single power law QLF with exponential density evolution confirms these results with a best fit bright-end slope of $\beta\approx-4.1$ for the full ELQS QLF sample. Our analysis further constrains the bright-end slope to be steeper than $\beta\leq-3.4$ at the $3\sigma$ level. Additionally, we perform broken double power law fits to the data to assess the possible bias introduced by a single power law description. This analysis corroborates the steep values for the bright-end slopes.
 While earlier studies \citep{Koo1988, Schmidt1995, Fan2001b, Richards2006} suggested a flattening of the bright-end slope $\beta$ towards higher redshifts all our analyses disfavor this scenario. In fact, our results at the intermediate redshift range rather encourage a consistent picture, in which the bright-end slope remains steep from the lowest redshifts \citep{Croom2009, Ross2013} up to the highest redshifts \citep{Jiang2008, Willott2010a, McGreer2013, Yang2016} with some room for modest evolution.
 
 \item We use an exponential density evolution model to analyze the redshift evolution of the quasar number density. The differential marginal distribution and the maximum likelihood fit consistently encourage an exponential decline with $\gamma\approx-0.4$. Other studies at lower and higher redshift \citep{Ross2013, Fan2001b, Yang2016} find a steeper decline of the quasar density towards higher redshift with $\gamma\sim-0.7$ to $-0.5$. However, the uncertainties on our maximum likelihood fit would allow for $\gamma\approx-0.5$ at the $1\sigma$ level.

 \item We combine the binned ELQS QLF with values from the \citet{Richards2006} and \citet{Ross2013} binned QLFs to calculate broken DPL fits over a larger magnitude range. Our best-fit results find the bright-end slope to be steep with values of $\beta\approx-4.6$ over $z=3.0-4.5$ and the break magnitude to be brighter by one magnitude compared to the previous literature. Only the recent re-analysis of a large combined sample of quasar surveys by \citet{Kulkarni2018} shows agreement with our results.
 We argue that the larger dynamic range in $M_{1450}$ probed by the ELQS survey at the bright-end was crucial to constrain the bright-end slope properly as previous studies \citep{Richards2006, Ross2013} did not sufficiently sample the population brighter than then break magnitude at these redshifts.

\end{enumerate}

\subsection*{Acknowledgements}
JTS, XF, IDM, and JY acknowledge support of US NSF Grant AST-1515115 and NASA ADAP Grant NNX17AF28G
 
JKK acknowledges financial support from the Danish Council for Independent Research (EU-FP7 under the Marie- Curie grant agreement no. 600207) with reference DFF- MOBILEX–5051-00115. 

The Cosmic Dawn Center is funded by the DNRF.

Y.-L. W. is grateful to the support from the Heising-Simons Foundation.

The work is in part based on observations made with the Nordic Optical Telescope, operated on the island of La Palma jointly by Denmark, Finland, Iceland, Norway, and Sweden, in the Spanish Observatorio del Roque de los Muchachos of the Instituto de Astrofisica de Canarias.

This publication makes use of data products from the Two Micron All Sky Survey, which is a joint project of the University of Massachusetts and the Infrared Processing and Analysis Center/California Institute of Technology, funded by the National Aeronautics and Space Administration and the National Science Foundation.

This publication makes use of data products from the Wide-field Infrared Survey Explorer, which is a joint project of the University of California, Los Angeles, and the Jet Propulsion Laboratory/California Institute of Technology, funded by the National Aeronautics and Space Administration.

Funding for the Sloan Digital Sky Survey IV has been provided by the Alfred P. Sloan Foundation, the U.S. Department of Energy Office of Science, and the Participating Institutions. SDSS acknowledges support and resources from the Center for High-Performance Computing at the University of Utah. The SDSS web site is \url{www.sdss.org}.

SDSS is managed by the Astrophysical Research Consortium for the Participating Institutions of the SDSS Collaboration including the Brazilian Participation Group, the Carnegie Institution for Science, Carnegie Mellon University, the Chilean Participation Group, the French Participation Group, Harvard-Smithsonian Center for Astrophysics, Instituto de Astrofísica de Canarias, The Johns Hopkins University, Kavli Institute for the Physics and Mathematics of the Universe (IPMU) / University of Tokyo, Lawrence Berkeley National Laboratory, Leibniz Institut f\"ur Astrophysik Potsdam (AIP), Max-Planck-Institut f\"ur Astronomie (MPIA Heidelberg), Max-Planck-Institut f\"ur Astrophysik (MPA Garching), Max-Planck-Institut f\"ur Extraterrestrische Physik (MPE), National Astronomical Observatories of China, New Mexico State University, New York University, University of Notre Dame, Observatório Nacional / MCTI, The Ohio State University, Pennsylvania State University, Shanghai Astronomical Observatory, United Kingdom Participation Group, Universidad Nacional Autónoma de México, University of Arizona, University of Colorado Boulder, University of Oxford, University of Portsmouth, University of Utah, University of Virginia, University of Washington, University of Wisconsin, Vanderbilt University, and Yale University.

This research has made use of the NASA/IPAC Extragalactic Database (NED) which is operated by the Jet Propulsion Laboratory, California Institute of Technology, under contract with the National Aeronautics and Space Administration.

This research made use of Astropy, a community-developed core Python package for Astronomy (Astropy Collaboration, 2013, \url{http://www.astropy.org}). In addition, python routines from scikit-learn\citep{scikit-learn}, SciPy \citep{scipy}, Matplotlib \citep{matplotlib}, lmfit \citep{lmfit2014}, and Pandas \citep{pandas} were used in the quasar selection, data analysis and creation of the figures.

\bibliographystyle{apj}

\appendix

\section{The full ELQS Quasar Catalog} \label{app_qso_catalog}
The ELQS quasar catalog is available as a machine readable table (csv format) on-line. It has 51 columns, detailed in Table\,\ref{tab_elqs_cat_cols}.

\begin{table*}[htp]
 \centering 
 \caption{Description of the full ELQS quasar catalog table}
 \label{tab_elqs_cat_cols}
 \begin{tabular}{cccp{8cm}}
  \tableline
  Column & Column Name & Unit & Description \\
  \tableline
  1 & wise\_designation &  - & Designation of the WISE AllWISE survey\\
  2 & sdss\_ra & deg & Right ascension from the SDSS DR13 \\
  3 & sdss\_dec & deg & Declination from the SDSS DR13 \\
  4 & sdss\_ra\_hms & hh:mm:ss.sss & Right ascension from the SDSS DR13 \\
  5 & sdss\_dec\_dms & dd:mm:ss.ss & Declination from the SDSS DR13 \\
  6 & wise\_ra & deg & Right ascension from the AllWISE \\
  7 & wise\_dec & deg & Declination from the AllWISE \\
  8 & reference & - & Reference to the quasar classification \\
  9 & reference\_z & - & Best redshift of the quasar according to the reference \\
  10 & M\_1450 & mag & Absolute magnitude at $1450\text{\AA}$ calculated using the k-correction determined in this work\\
  11 & sel\_prob & - & Selection probability according to our completeness calculation \\
  12-21 & [survey]\_mag\_[band] & mag & Dereddened AB magnitudes of the SDSS ugriz, 2MASS jh$\rm{k}_{\rm{s}}$ and WISE W1W2 bands (surveys = [SDSS,TMASS,WISE]; bands = [u,g,r,i,z],[j,h,k],[w1,w2]). It should be noted that all SDSS magnitudes are PSF magnitudes in the SDSS Asinh magnitude system.\\
  22-31 & [survey]\_magerr\_[band] & mag & $1\sigma$ errors on the AB magnitudes. \\
  32 & E\_BV & mag & E(B-V) \\
  33 & extinction\_i & mag & Extinction in the SDSS i-band\\ 
  34 & FIRST\_match & True/False & Boolean to indicate successful matches with the FIRST catalog \\
  35 & FIRST\_distance & arcsec & Distance of the FIRST source relative to the SDSS position \\
  36 & FIRST\_peak\_flux\_mJy/bm & mJy/bm & FIRST peak flux\\
  37 & FIRST\_RMS\_mJy/bm & mJy/bm & RMS error on the FIRST flux\\
  38 & GALEX\_match & True/NaN &  Boolean to indicate successful matches with the GALEX GR6/7 catalog\\
  39 & GALEX\_distance & arcsec & Distance of the GALEX GR6/7 match relative to the SDSS position \\
  40 & GALEX\_nuv\_mag & mag & GALEX near-UV flux in magnitudes \\
  41 & GALEX\_nuv\_magErr & mag & Error on the GALEX near-UV flux \\
  42 & GALEX\_fuv\_mag & mag & GALEX far-UV flux in magnitudes \\
  43 & GALEX\_fuv\_magErr & mag & Error on the GALEX far-UV flux \\
  44 & TRXS\_match & True/False & Boolean to indicate successful matches to the ROSAT 2RXS AllWISE counterparts \\
  45 & TRXS\_distance & arcsec & Match distance between the ELQS AllWISE position to the ROSAT 2RXS AllWISE position. The distance values are often 0 or otherwise extremely small, because the positions match to numerical accuracy.\\
  46 & TRXS\_match\_flag & - & A flag indicating the most probable AllWISE ROSAT 2RXS cross-match with 1. This is the case for all matched objects.\\
  47 & TRXS\_2RXS\_SRC\_FLUX & $\rm{erg}\,\rm{cm}^{-2}\,\rm{s}^{-1}$ &  2RXS flux \\
  48 & TRXS\_2RXS\_SRC\_FLUX\_ERR &  $\rm{erg}\,\rm{cm}^{-2}\,\rm{s}^{-1}$ & 2RXS flux error\\
  49 & BAL\_VI & 1/0/-1 & A flag indicating whether the object is visually identified as a broad absorption line (BAL) quasar ($1=$ BAL quasars, $0=$ quasars, $-1=$ no visual classification).\\
  50 & qlf\_sample & True/False & Boolean to indicate whether the quasar is included in the estimation of the quasar luminosity function (Section\,\ref{sec_QLF}).\\
  \tableline
 \end{tabular}

\end{table*}
\clearpage
\section{Discovery Spectra of the ELQS-S Quasars}

\begin{figure*}[htb]
 \centering
 \includegraphics[width=0.9\textwidth]{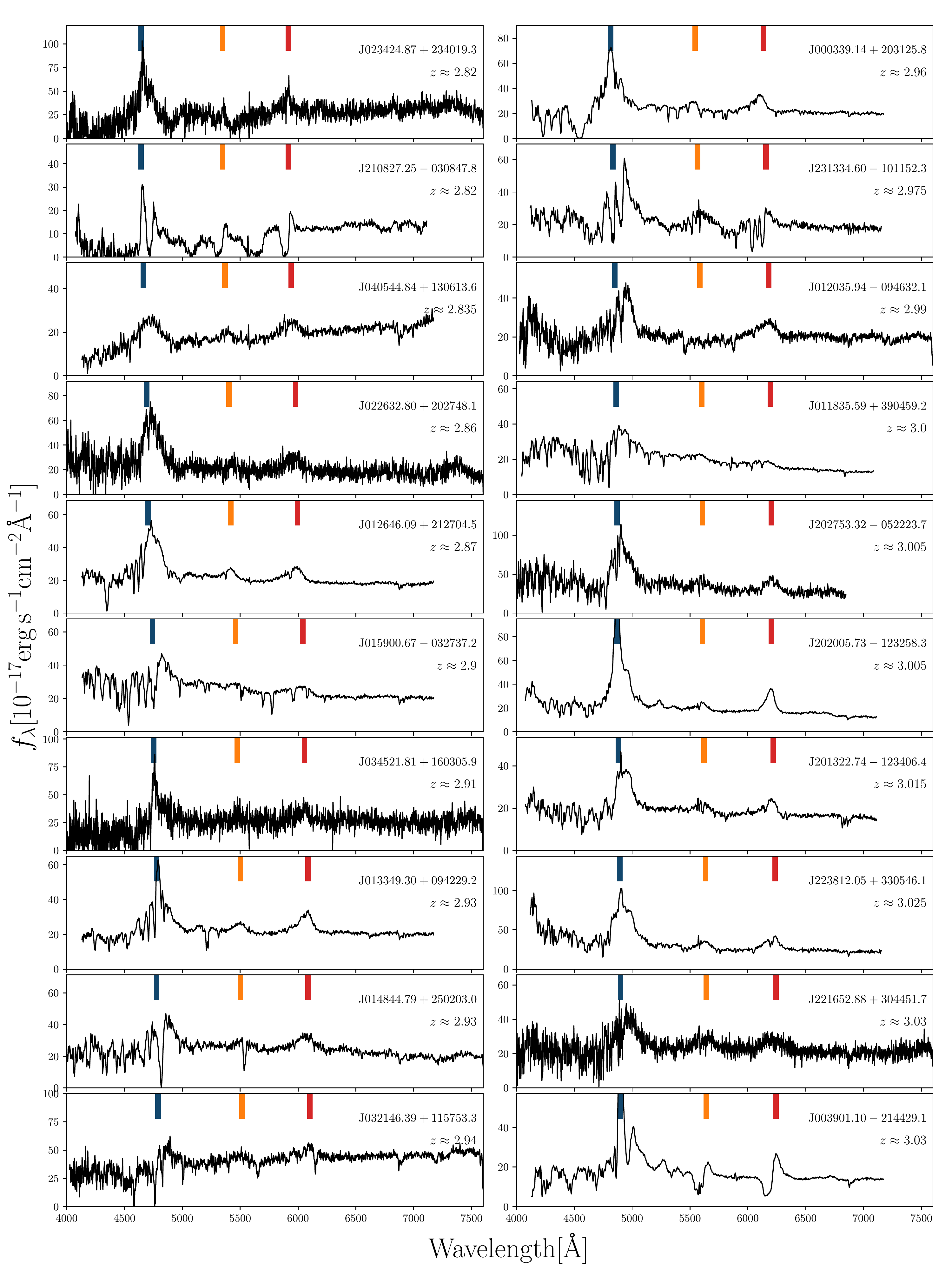}
 \caption{The discovery spectra of the newly identified ELQS-S quasars sorted by spectroscopic redshift. The  dark blue, orange and red bars denote the center positions of the broad $\rm{Ly}\alpha$, \ion{Si}{4} and \ion{C}{4} emission lines according to the spectroscopic redshift.}
 \label{fig_newqso_spectra}
\ContinuedFloat
\end{figure*}
 
\begin{figure*}[htb]
 \centering
 \includegraphics[width=0.9\textwidth]{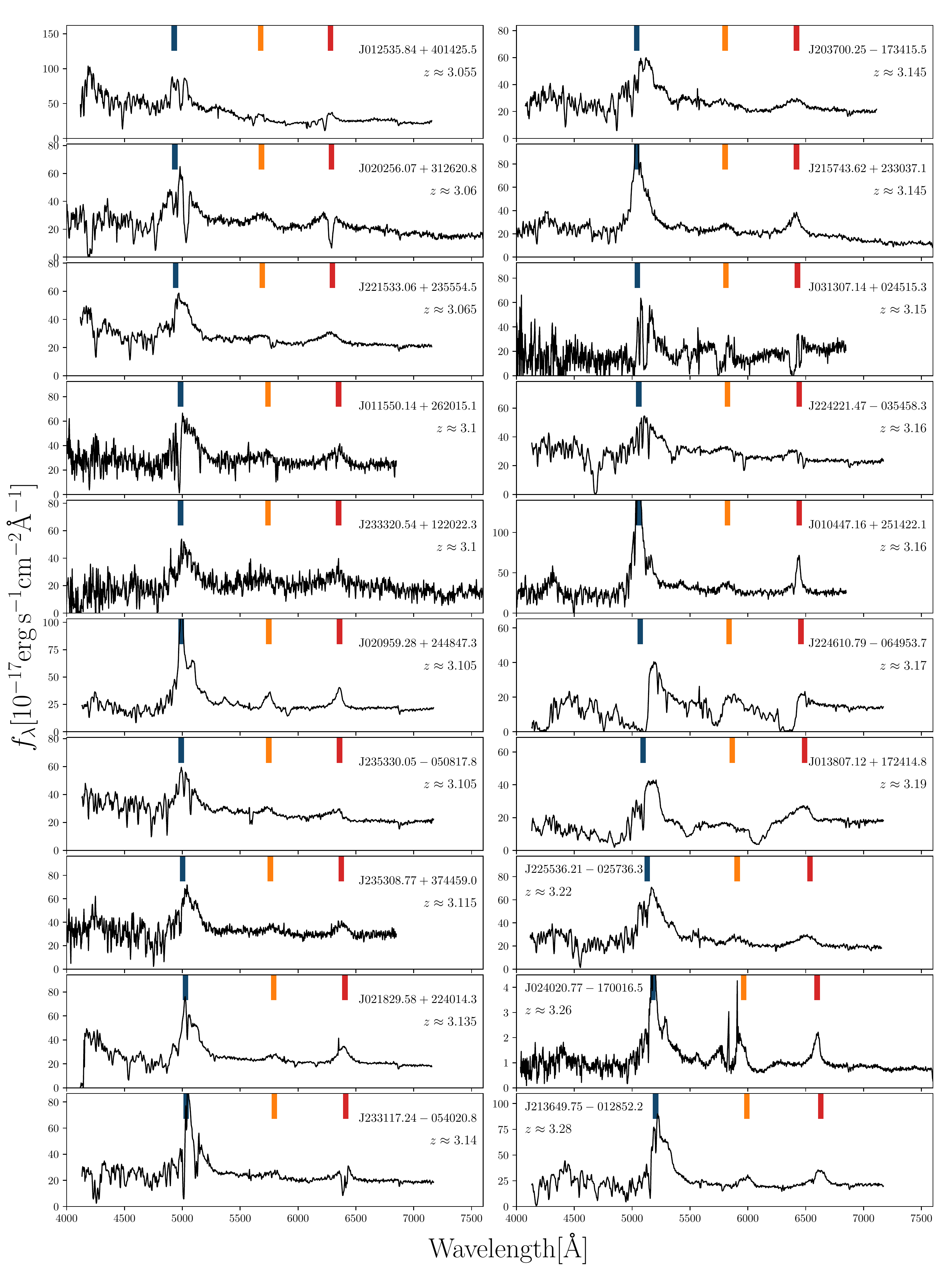}
 \caption[]{(continued)}
\ContinuedFloat
\end{figure*} 

\begin{figure*}[htb] 
 \centering
 \includegraphics[width=0.9\textwidth]{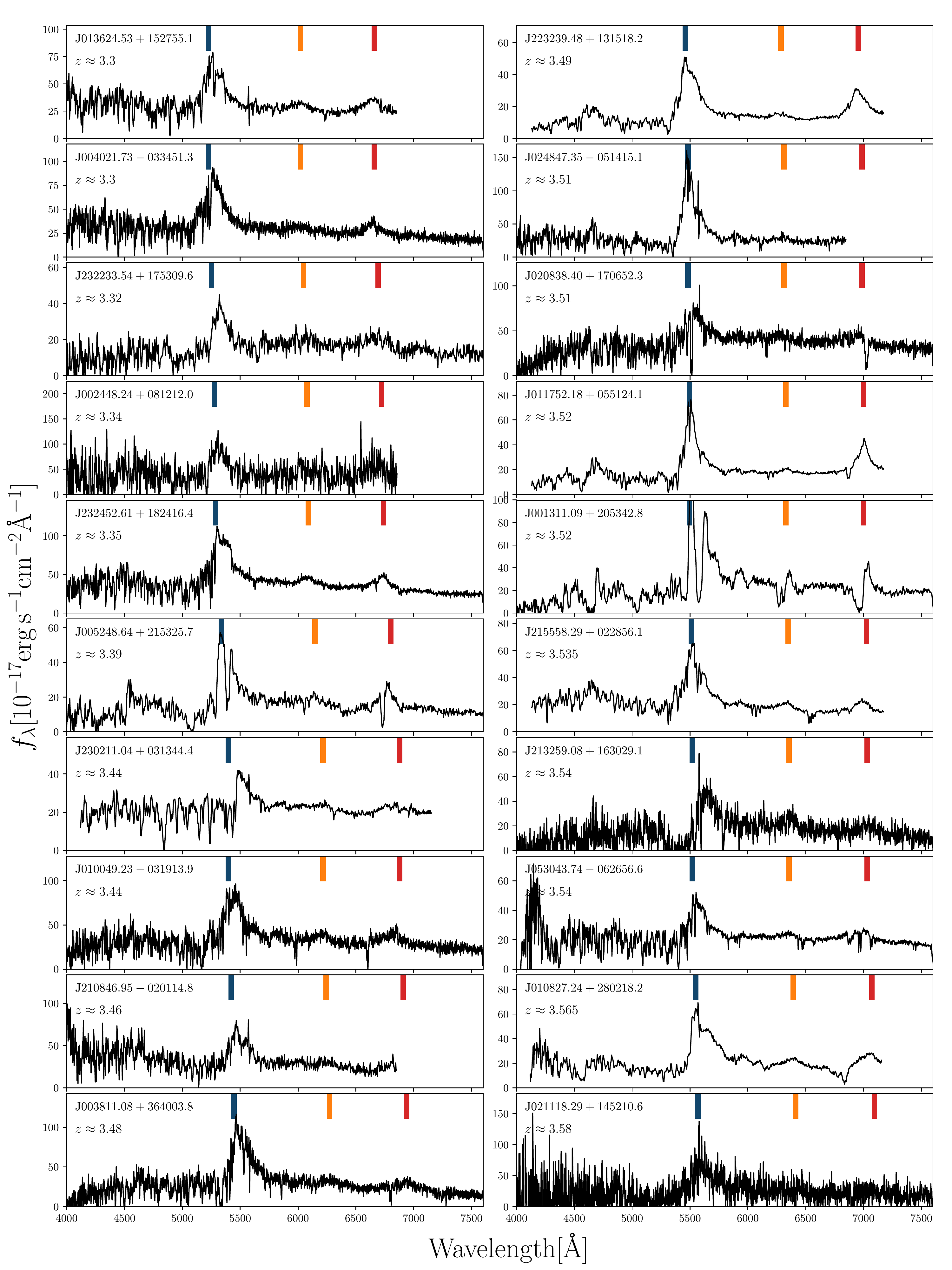}
 \caption[]{(continued)}
 \ContinuedFloat
\end{figure*} 

\begin{figure*}[htb] 
 \centering
 \includegraphics[width=0.9\textwidth]{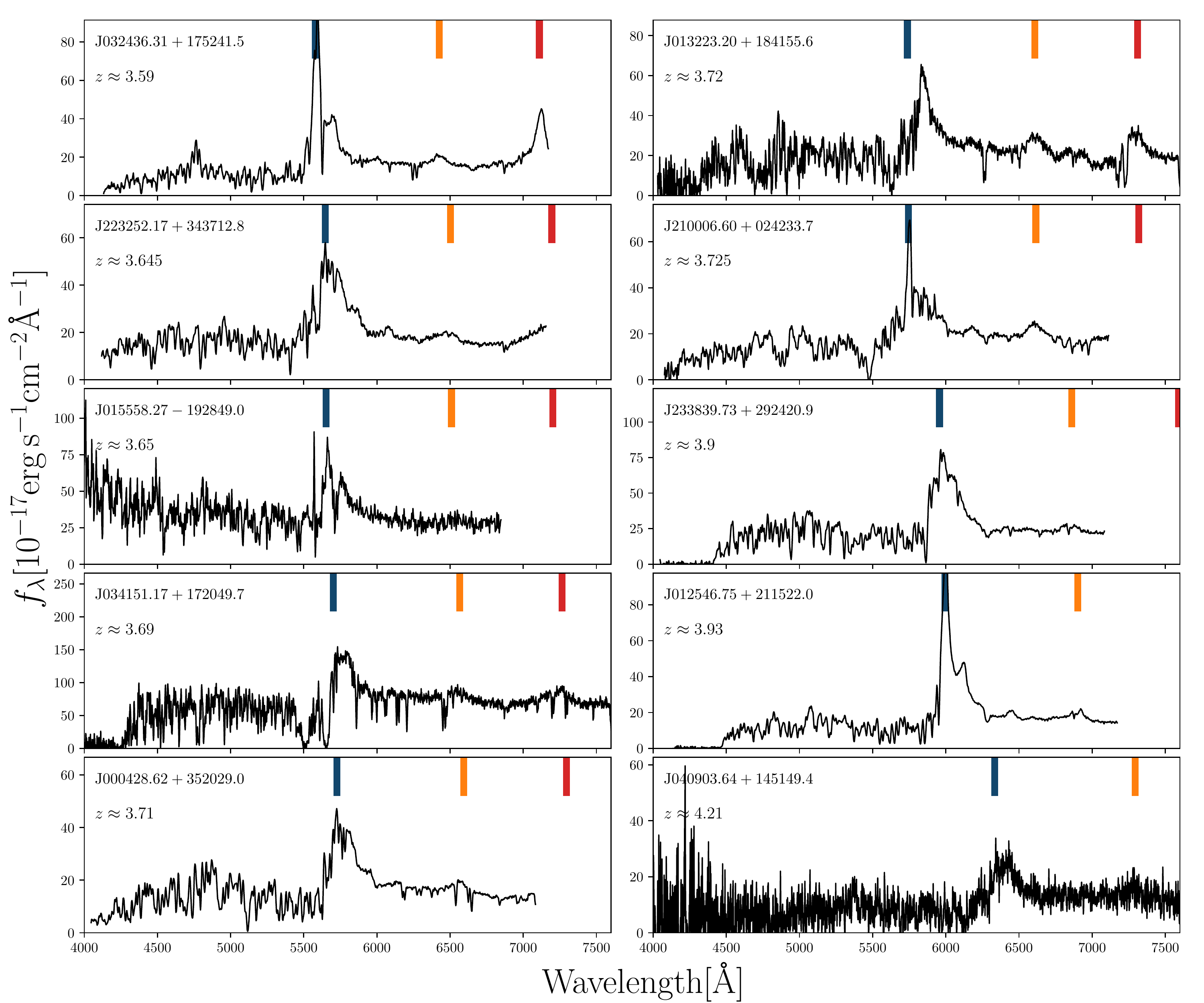}
 \caption[]{(continued)}
\end{figure*} 

\clearpage
\section{Discovery spectra of primary ELQS quasar candidates with $z<2.8$}

%

\begin{table*}[htb]
\centering
 \caption{Newly discovered quasars at $z<2.8$ in the ELQS sample}
 \label{tab_elqs_lowz_newqsos}
 \begin{tabular}{cccccccc}
  \tableline
  \tableline
 R.A.(J2000) & Decl.(J2000) & $m_{\rm{i}}$ &  $M_{1450}$ & Spectroscopic & near UV\tablenotemark{b} & far UV\tablenotemark{b}  & Notes \\
 
 [hh:mm:ss.sss] & [dd:mm:ss.ss] & [mag]  & [mag] &  Redshift & [mag] & [mag] &  \\
  \tableline
 00:48:49.500 & +38:31:16.73 & $17.77\pm0.03$ & - & ? & - &  - & 171020 \\
 01:16:37.968 & +22:11:47.31 & $17.40\pm0.02$ & -27.66 & 2.790 & - &  - & 170825 \\
 02:27:42.939 & -17:31:21.54 & $17.58\pm0.02$ & -27.12 & 2.300 & - &  - & 171021 \\
 09:06:19.161 & +39:29:32.35 & $17.95\pm0.02$ & -27.09 & 2.720 & $21.00\pm0.33$ &  - & 170517 \\
 09:31:33.416 & +17:20:48.64 & $16.79\pm0.01$ & -28.27 & 2.755 & - &  - & 161122 \\
 12:05:04.556 & +02:57:19.20 & $17.75\pm0.01$ & -27.17 & 2.495 & $22.44\pm0.19$ &  - & 160311 \\
 12:11:19.771 & +30:41:33.25 & $17.75\pm0.02$ & -27.31 & 2.780 & - &  - & 170518 \\
 13:26:25.921 & +15:22:16.49 & $17.62\pm0.02$ & -27.43 & 2.765 & - &  - & 170405 \\
 13:55:33.171 & +56:38:32.25 & $17.90\pm0.02$ & - & ? & - &  - & 170503 \\
 13:57:43.325 & -06:00:47.14 & $17.51\pm0.01$ & -27.47 & 2.620 & - &  - & 170405 \\
 13:59:56.032 & +06:14:30.19 & $17.03\pm0.02$ & -26.66 & 1.515 & $22.52\pm0.18$ &  - & 170404 \\
 14:47:50.137 & +32:03:50.29 & $17.19\pm0.01$ & - & ? & - &  - & 170503 \\
 15:05:51.111 & +05:19:57.32 & $17.43\pm0.01$ & - & ? & - &  - & 170504 \\
 15:19:06.817 & +26:43:26.29 & $17.96\pm0.01$ & -27.10 & 2.770 & - &  - & 150509 \\
 15:59:29.631 & +34:13:16.12 & $17.84\pm0.02$ & -27.21 & 2.740 & - &  - & 170406 \\
 16:58:20.175 & +15:47:58.97 & $17.87\pm0.01$ & -27.16 & 2.720 & - &  - & 170405 \\
 17:07:03.862 & +20:25:39.27 & $18.00\pm0.04$ & -27.05 & 2.740 & - &  - & 170419 \\
 17:15:11.188 & +62:55:26.10 & $17.58\pm0.02$ & -26.14 & 1.580 & - &  - & 170503 \\

\tableline
 \end{tabular}
\tablenotetext{1}{These objects were also independently discovered by Yang et al.}
\tablenotetext{2}{The near and far UV magnitudes were obtained from cross-matches within $2\farcs0$ to the GALEX GR6/7 data release.}
\end{table*}

\begin{figure*}[htb]
 \centering
 \includegraphics[width=0.9\textwidth]{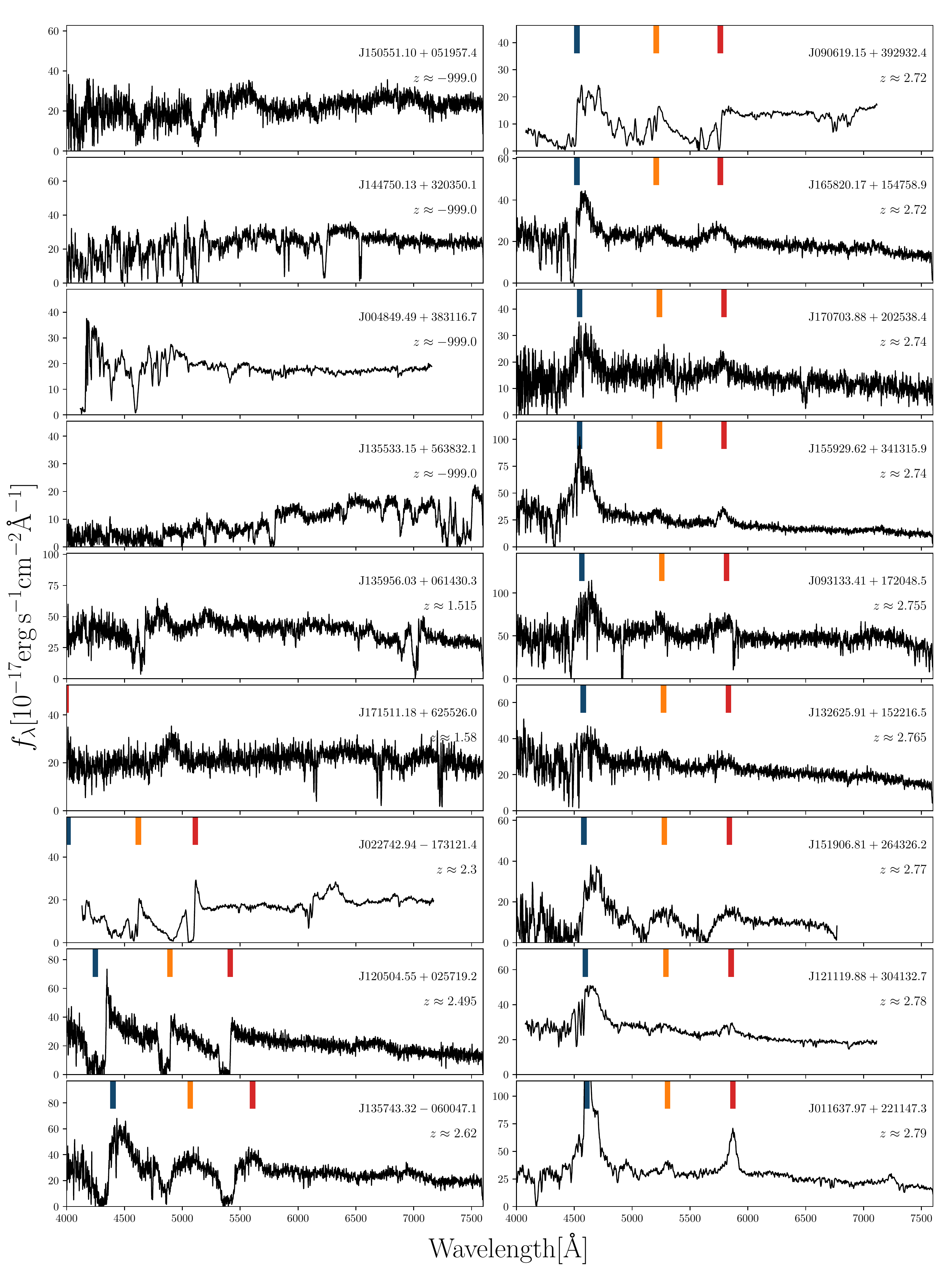}
 \caption{The discovery spectra of primary ELQS candidates at $z<2.8$. The  dark blue, orange and red bars denote the center positions of the broad $\rm{Ly}\alpha$, \ion{Si}{4} and \ion{C}{4} emission lines according to the spectroscopic redshift.}
 \label{fig_lowz_spectra}
\end{figure*}

\clearpage
\section{Additional Quasars discovered as part of this project} \label{app_addquasars}

\begin{table*}[htb]
\centering
 \caption{Newly discovered quasars not part of the ELQS primary candidate catalog}
 \label{tab_elqs_sec_newqsos}
 \begin{tabular}{cccccc}
  \tableline
  \tableline
 R.A.(J2000) & Decl.(J2000) & $m_{\rm{i}}$ &  $M_{1450}$ & Spectroscopic  & Notes \\
 
 [hh:mm:ss.sss] & [dd:mm:ss.ss] & [mag]  & [mag] &  Redshift &  \\
\tableline
00:55:42.504 & -06:14:11.99 & $18.43\pm0.02$ & -27.06 & 3.390 & 151201 \\
01:26:30.629 & +08:32:25.35 & $17.34\pm0.01$ & -27.71 & 2.770 & 161115 \\
02:19:35.796 & +19:02:55.17 & $17.21\pm0.01$ & -27.83 & 2.730 & 161122 \\
02:41:23.499 & -07:05:12.36 & $17.38\pm0.02$ & -27.67 & 2.770 & 161123 \\
03:06:56.870 & -07:54:28.60 & $18.20\pm0.02$ & - & ? & 151012 \\
04:41:55.219 & -01:03:49.98 & $17.22\pm0.01$ & - & ? & 161123 \\
05:15:54.937 & +01:22:42.59 & $18.17\pm0.01$ & -26.92 & 2.820 & 161123 \\
08:06:36.114 & +48:40:26.69 & $17.30\pm0.02$ & -27.70 & 2.680 & 161218 \\
08:27:46.205 & +82:00:49.95 & $18.16\pm0.02$ & -27.05 & 2.975 & 170405 \\
09:12:23.780 & +12:44:08.14 & $18.19\pm0.02$ & -27.02 & 2.995 & 170518 \\
09:46:32.282 & +66:32:24.61 & $18.36\pm0.02$ & -26.85 & 2.995 & 170517 \\
11:04:42.076 & +45:46:43.26 & $18.21\pm0.02$ & -27.34 & 3.610 & 170419 \\
11:09:45.306 & +13:57:22.20 & $18.14\pm0.02$ & -27.38 & 3.505 & 170419 \\
11:13:32.451 & -03:09:14.07 & $18.05\pm0.01$ & -27.60 & 3.740 & 170418 \\
11:28:30.664 & +75:15:20.94 & $18.21\pm0.02$ & -26.88 & 2.810 & 170406 \\
11:29:47.663 & +41:06:57.05 & $18.18\pm0.02$ & -27.33 & 3.480 & 150508 \\
11:38:40.635 & +34:35:54.25 & $18.01\pm0.02$ & -27.36 & 3.230 & 170418 \\
11:40:05.747 & +71:53:16.17 & $18.27\pm0.03$ & -27.19 & 3.365 & 170405 \\
11:48:11.638 & -01:40:24.55 & $18.28\pm0.02$ & -26.93 & 3.010 & 170419 \\
12:01:15.165 & +30:13:58.47 & $17.97\pm0.02$ & -26.90 & 2.460 & 160312 \\
12:01:16.305 & +26:16:11.89 & $18.07\pm0.02$ & - & ? & 170417 \\
12:21:53.197 & +23:53:24.39 & $18.09\pm0.03$ & -27.67 & 3.930 & 170518 \\
12:43:40.542 & +24:01:42.14 & $18.12\pm0.02$ & -26.97 & 2.830 & 160312 \\
13:14:17.573 & +38:45:17.13 & $18.11\pm0.02$ & -26.94 & 2.760 & 170406 \\
13:18:43.193 & +38:23:34.36 & $18.11\pm0.02$ & -27.66 & 3.970 & 160312 \\
13:19:23.907 & -00:26:20.55 & $18.21\pm0.01$ & -26.98 & 2.960 & 170417 \\
13:58:58.050 & -03:29:08.33 & $18.41\pm0.02$ & - & ? & 150422 \\
14:23:57.190 & -18:53:47.33 & $18.36\pm0.02$ & -25.81 & 1.930 & 170518 \\
14:59:01.000 & -02:51:05.81 & $18.01\pm0.01$ & -27.43 & 3.320 & 160310 \\
15:20:44.943 & +38:25:12.27 & $18.07\pm0.01$ & -27.11 & 2.930 & 170406 \\
15:36:31.102 & +34:47:59.58 & $18.35\pm0.02$ & -27.14 & 3.430 & 160312 \\
15:46:19.728 & +36:10:40.41 & $18.21\pm0.02$ & -26.98 & 2.950 & 170406 \\
15:46:37.131 & -02:31:07.62 & $18.14\pm0.02$ & -27.68 & 4.050 & 170418\tablenotemark{a} \\
15:53:40.946 & +06:57:38.44 & $18.05\pm0.02$ & -27.51 & 3.600 & 170404 \\
16:04:52.284 & +38:47:55.06 & $17.40\pm0.02$ & -27.66 & 2.800 & 160311 \\
16:21:33.452 & +43:46:28.41 & $17.72\pm0.01$ & -27.32 & 2.740 & 160313\tablenotemark{a} \\
16:34:00.284 & +64:08:22.10 & $18.31\pm0.02$ & -27.25 & 3.630 & 170406 \\
16:35:25.313 & +38:14:29.42 & $18.25\pm0.01$ & -26.83 & 2.790 & 170419 \\
16:40:03.561 & +53:26:33.67 & $18.31\pm0.01$ & -27.02 & 3.165 & 170419 \\
16:44:05.048 & +53:42:49.90 & $18.39\pm0.02$ & -27.12 & 3.510 & 170419 \\
16:48:52.294 & +52:09:51.37 & $18.13\pm0.01$ & -27.47 & 3.685 & 170406\tablenotemark{a} \\
16:58:28.042 & +50:23:07.98 & $18.09\pm0.01$ & -27.17 & 3.080 & 170406 \\
17:08:44.725 & +28:27:30.49 & $17.60\pm0.01$ & -27.54 & 2.890 & 160312 \\
17:17:21.347 & +42:24:28.12 & $18.17\pm0.02$ & -27.35 & 3.490 & 170419 \\
17:45:48.100 & +46:33:45.71 & $18.03\pm0.03$ & -27.19 & 3.010 & 160313 \\
20:37:05.975 & -13:45:57.50 & $18.07\pm0.03$ & -27.30 & 3.230 & 170518 \\
20:59:07.574 & +08:36:44.19 & $18.26\pm0.02$ & -27.07 & 3.170 & 170518 \\
22:11:24.148 & +25:43:27.16 & $16.94\pm0.03$ & -28.14 & 2.805 & 161115 \\
22:51:59.484 & +17:28:44.64 & $17.22\pm0.01$ & - & ? & 171110 \\
22:56:43.450 & -02:46:32.83 & $17.65\pm0.01$ & -27.16 & 2.400 & 151011 \\
23:58:28.105 & +24:07:46.70 & $18.01\pm0.01$ & -27.90 & 4.180 & 161123 \\

\tableline
 \end{tabular}
\tablenotetext{1}{These objects were also independently discovered by Yang et al.}
\end{table*}
{
\begin{figure*}[htb]
 \centering
 \includegraphics[width=0.9\textwidth]{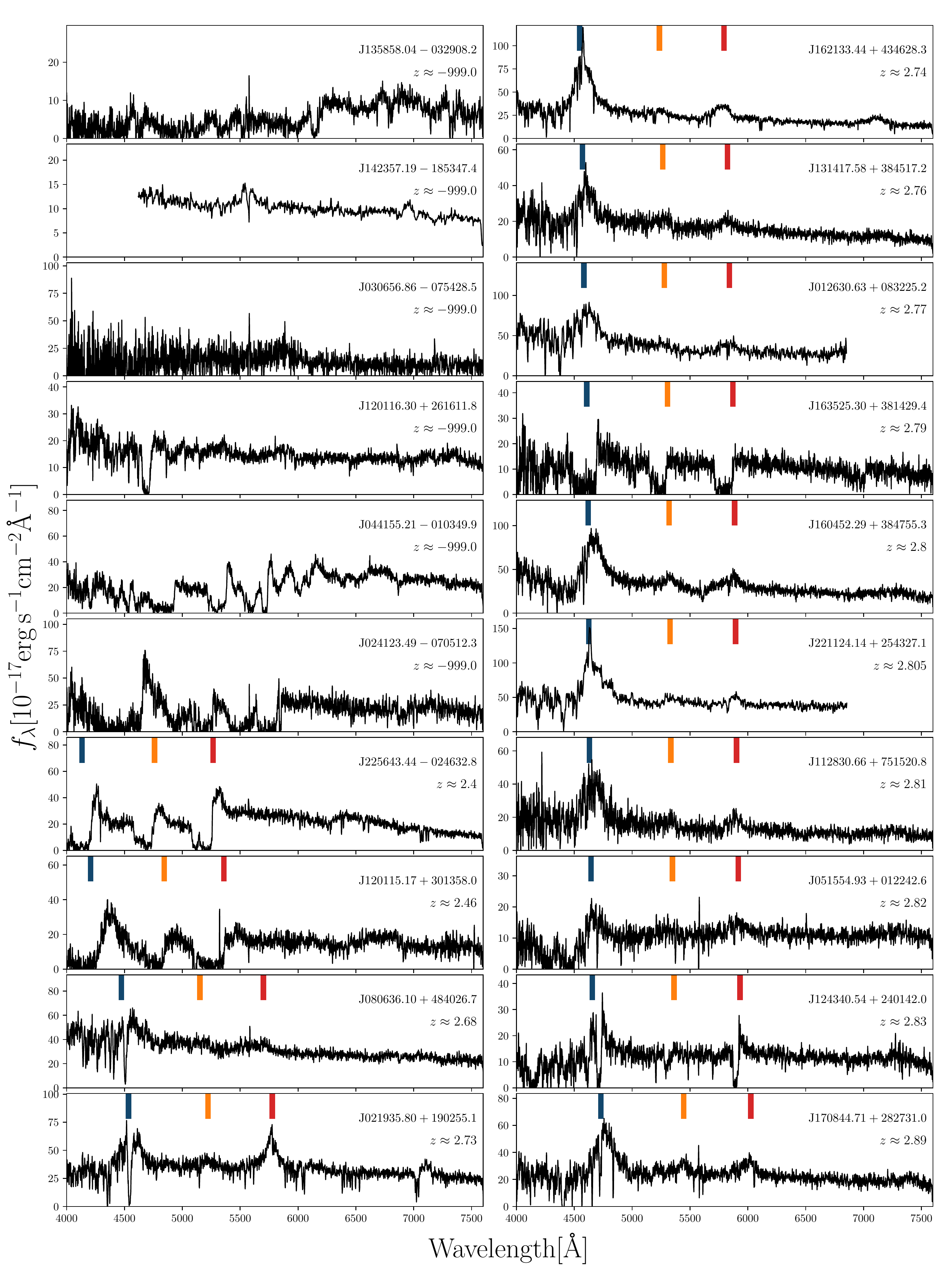}
 \caption{The discovery spectra of additional quasars discovered during the first stages of the ELQS survey. The majority of these objects are fainter than $i=18.0$ and therefore not included in the final ELQS candidate catalog. The  dark blue, orange and red bars denote the center positions of the broad $\rm{Ly}\alpha$, \ion{Si}{4} and \ion{C}{4} emission lines according to the spectroscopic redshift.}
 \label{fig_sec_spectra}
\ContinuedFloat
\end{figure*}
 
\begin{figure*}[htb] 
 \centering
 \includegraphics[width=0.9\textwidth]{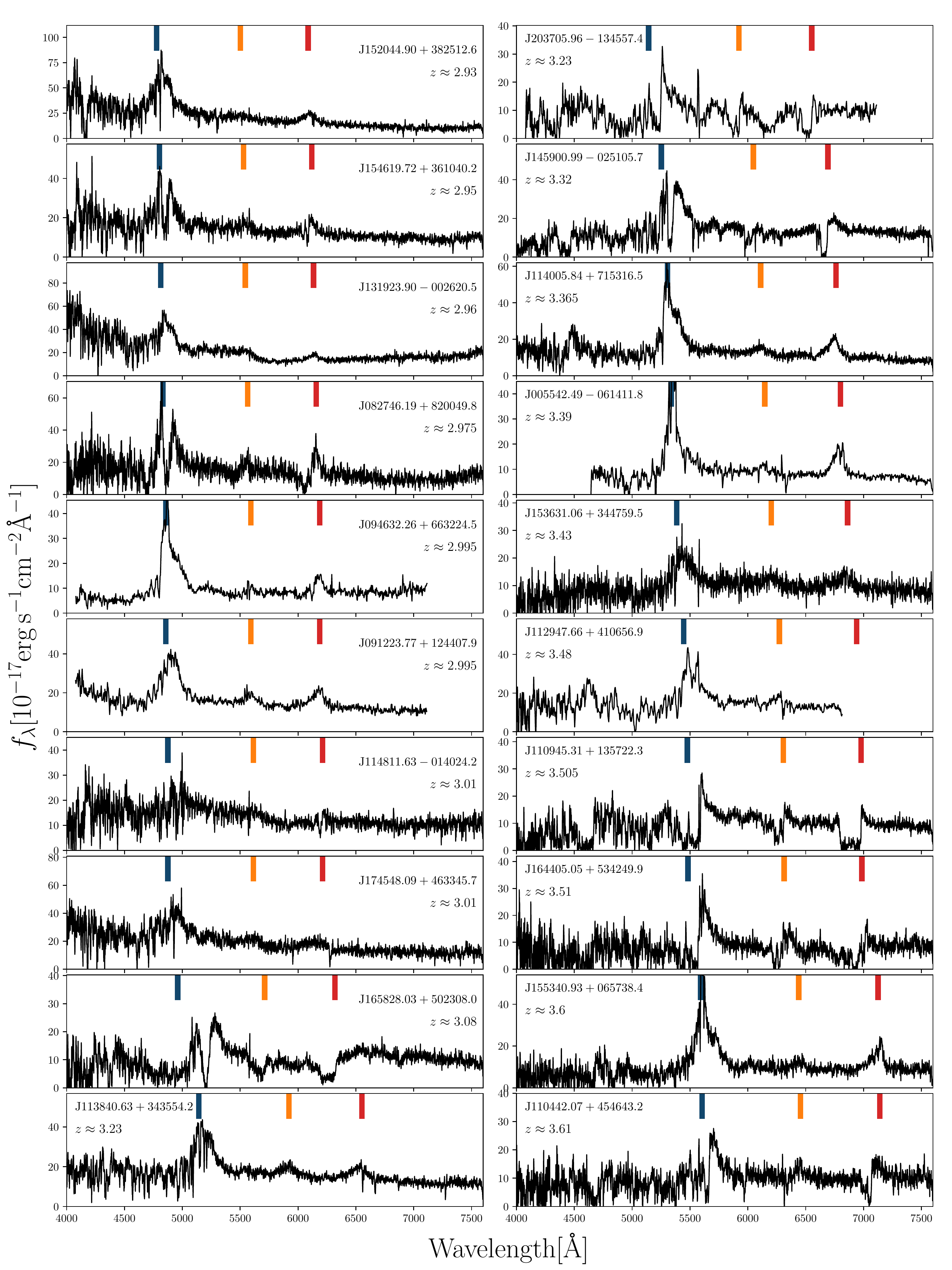}
 \caption[]{(continued)}
\ContinuedFloat
\end{figure*} 

\begin{figure*}[htb] 
 \centering
 \includegraphics[width=0.9\textwidth]{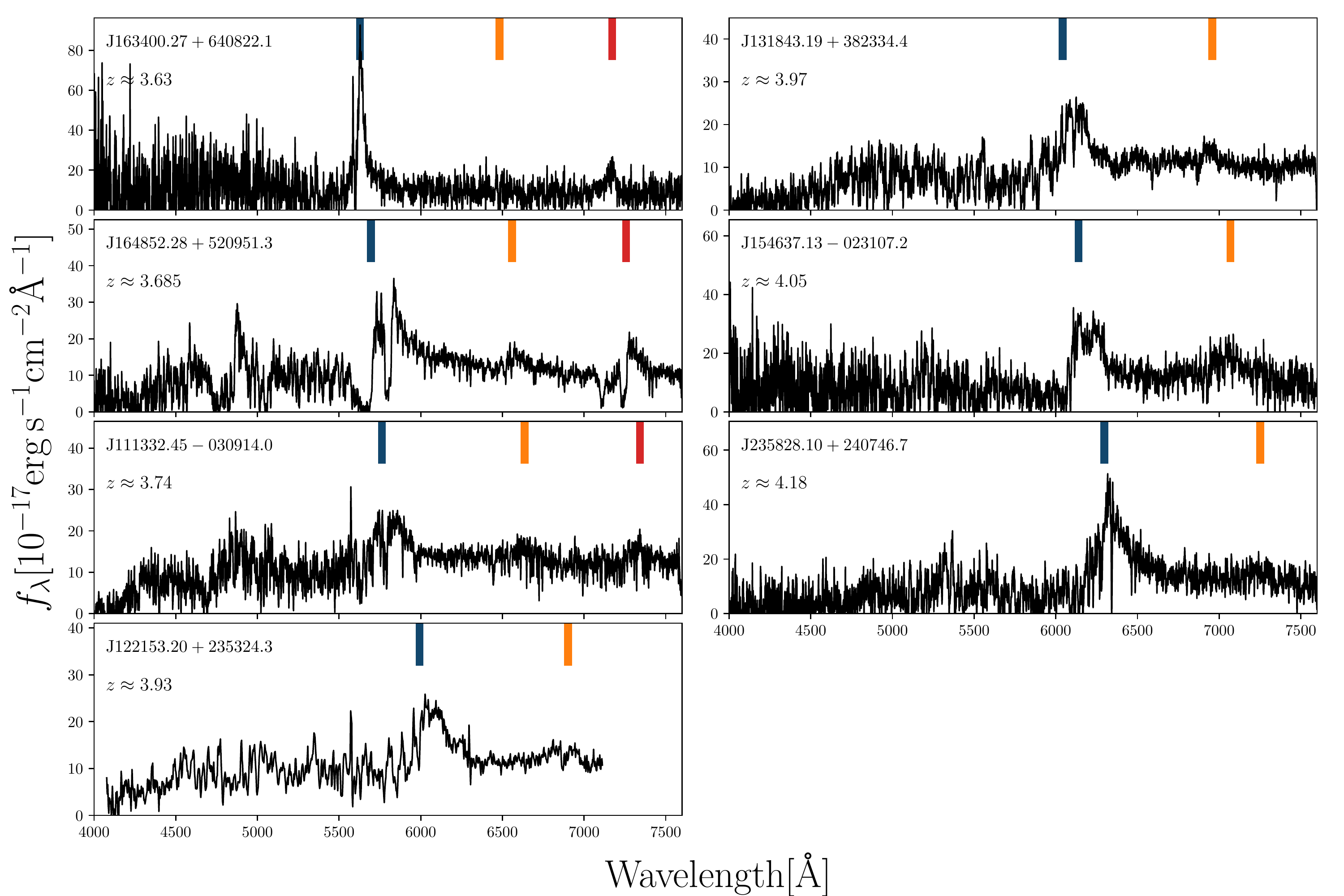}
 \caption[]{(continued)}
\end{figure*} 
}
\end{document}